\documentclass[aps,pra,showpacs,reprint]{revtex4-2}
\usepackage{lineno}
\usepackage[b]{esvect}

\usepackage{graphicx}
\usepackage[normalem]{ulem}

\usepackage{stackengine}
\usepackage{amsmath}
\usepackage{xcolor}

\stackMath 

\newcommand{\lowarrow}[1]{\stackon[-0.5pt]{#1}{\leftrightarrow}}

\let\originaleqref\eqref
\renewcommand{\eqref}[1]{Eq.~\originaleqref{#1}}

\usepackage{amsmath,amssymb} 

\begin{document}

\title{Correlation geometry and topology of 
structured optical beams}

\author{Jyrki Laatikainen}
\affiliation{Department of Physics, University of Miami, 1320 Campo Sano Avenue, Coral Gables, FL, 33146, USA}

\author{Olga Korotkova} \email{korotkova@physics.miami.edu}
\affiliation{Department of Physics, University of Miami, 1320 Campo Sano Avenue, Coral Gables, FL, 33146, USA}

\date{\today}
\begin{abstract}
Correlation geometry and topology in a random, scalar beam carrying Orbital Angular Momentum (OAM) in $L$ modes are shown to be linked to the real and imaginary parts, respectively, of its orbitalization matrix (OM). The OM is obtained by filtering the cross-spectral density in the polar Fourier basis at a given cross-section and radius for each pair of OAM indices. The  symmetric real part of the OM has the diagonal canonical form, specifying the ellipsoid-like correlation geometry of the beam. The skew-symmetric imaginary part of the OM has the Darboux canonical form, enabling decomposition into $N\leq L/2$ mutually orthogonal circulation states and $L-2N$ trivial circulation-free states,
thereby defining the $N$-dimensional orbitalization torus associated with the correlation topology of the beam. The canonical forms are then used to define the degrees of linear and circular orbital anisotropy and stability, and the degrees of linear and circular orbitalization, in analogy with corresponding quantities in 2D and 3D polarization theory. These results demonstrate the emergence of fundamentally new correlation structures in random multimode OAM-carrying light, not accessible in lower-dimensional systems and absent in the deterministic limit.
\end{abstract}

\maketitle

\section{Introduction}

Beam-like, random, vectorial optical fields are locally characterized by a $2\times 2$ polarization matrix defining a single elliptical polarization state which can degenerate to circular or linear, or vanish in the case of trivial degree of polarization \cite{BW}. In particular, circular states and helicity are related to the imaginary part of the polarization matrix. For general, non-beam like fields characterized by $3\times 3$ polarization matrices \cite{Brosseau1998,Gil}, the polarization plane may fluctuate, and different approaches have been developed to characterize the resulting 3D polarization structure \cite{Setala2002,Dennis2004,Ellis2004,Ellis2005degree,Luis2005,Gil2017,Gil2018}.

On the other hand, even in scalar approximation, light beams can be constructed as superpositions or correlations of a finite number $L$ of structured beams, i.e., vortex modes 
each carrying Orbital Angular Momentum (OAM) with topological charge $l$ \cite{SO}. In a deterministic case, such a construction gives rise to an $L$-dimensional ($L$D) isotropic oscillator in the basis $e^{il\phi}$ where $\phi$ is a polar angle. Unlike the point-based, real-space 2D or 3D polarization oscillator, the OAM oscillator is defined at a radius and belongs to the polar Fourier space of the beam, accessible with an appropriate OAM-mode-filtering system \cite{measure}. 
As any isotropic oscillator, OAM oscillator traces an elliptic trajectory, Orbitalization Ellipse (OE), in its planar subspace, regardless of $L$ \cite{OE}. The ellipse is also present in a random structured light beam, characterized by $L\times L$ Orbitalization Matrix (OM), and 
its size is determined by the degree of orbitalization \cite{OErandom}. 

The aim of this paper is to decompose the OM into its real and imaginary parts and, by finding their canonical forms, to describe their contributions to the correlation structure of random multimode OAM-carrying beams through the associated geometric and topological manifolds and correlation measures. This analysis hence provides thorough comprehension of the correlation structure in such beams and establishes, for an arbitrary number of OAM modes, a framework analogous to the polarization description of vectorial optical fields. The paper is organized as follows. Section 2 presents the general properties of the real and imaginary parts of the OM and their connections to the OAM flux density. Section 3 establishes the eigenmodes of the real and imaginary portions of the OM, and shows that these parts encode the linear and circular orbitalization information of the beam, respectively. 
In particular, Sec. 3A utilizes the real part of the OM to define the orbitalization ellipsoid of the beam. Further, Sec. 3B derives the Darboux form associated with the imaginary part of the OM and, based on it, defines the circulation state of the beam and the orbitalization torus. Section 3C then relates the eigenmodes of the OM to those of its real and imaginary parts. Section 4 analyzes the complete eigenmode decomposition of the OM and introduces the degrees of linear and circular orbital anisotropy and stability. These quantify, respectively, the anisotropy among the canonical real-part mode weights, the overall strength of the circulation-producing imaginary-part correlations, and the concentration of the real-part correlation structure within its principal 2D subspace. In addition, the connection of these measures to the degree of orbital anisotropy of the beam is established. Section 5 focuses on the fully orbitalized portion of the beam and defines the degrees of linear and circular orbitalization associated with it. Finally, Sec. 6 contains a brief summary of the results.

\section{Real and imaginary parts of orbitalization matrix}
Consider a wide-sense stationary, scalar light beam whose spatial coherence is described by the cross-spectral density function $W(\mathbf{r}_1,\mathbf{r}_2,\omega)=\langle E^\ast (\mathbf{r}_1,\omega)E(\mathbf{r}_2,\omega)\rangle$ \cite{Mandel}. Here 
$E(\textbf{r},\omega)$ are monochromatic realizations at angular frequency $\omega$ of the statistical ensemble representing the field, defined at a point with position vector $\textbf{r}=\rho\hat{\rho}+z\hat{z}+\phi\hat{\phi}$, resolved in cylindrical system with $\rho$, $z$, and $\phi$ being radial, axial, and azimuthal coordinates, respectively. In addition, the angle brackets and asterisk stand for ensemble averaging and complex conjugate, respectively. Denoting the axially symmetric portion of the coordinate system by $\pmb{\xi}=\rho\hat{\rho}+z\hat{z}$ we expand the field into polar Fourier (OAM) series as \cite{Gori2001,COAM} 
\begin{align}\label{SpFs}
E(\textbf{r},\omega)&=\sum\limits_{l\in\mathcal{L} }E_l(\pmb{\xi})e^{il\phi}, \\ E_l(\pmb{\xi})&=\frac{1}{2\pi}\int_0^{2\pi}E(\mathbf{r},\omega)e^{-il\phi}d\phi, \label{SpFs-2}
\end{align}
where the frequency dependence in $E_l(\pmb{\xi})$ has been omitted for brevity. For practical purposes we assume that a finite number 
$L$ of components $E_l(\pmb{\xi})$
is invoked, i.e., OAM index $l\in\mathcal{L} \subset \mathbb{Z}.$ Components $E_l(\pmb{\xi})$ can be grouped into an $L\times 1$ vector $\vec{E}(\pmb{\xi})=[E_{l}(\pmb{\xi})]$, being the counterpart of a Jones vector \cite{BW} in the OAM space.

In a fixed transverse cross-section, the single-radius, pair-OAM mode, second-order correlations in the beam are described by $L\times L$ Hermitian, positive semi-definite (PSD) OM \cite{OErandom}
\begin{equation}\label{OM}
\lowarrow{O}(\pmb{\xi})= 
\langle \vec{E}^* (\pmb{\xi}) \vec{E}^T (\pmb{\xi})\rangle,
\end{equation}
with elements 
\begin{equation}
    O_{lm}(\pmb{\xi})=\langle E_l^*(\pmb{\xi})E_m(\pmb{\xi})\rangle=|O_{lm}(\pmb{\xi})|e^{i\varphi_{lm}(\pmb{\xi})},
\end{equation}
where $l,m \in\mathcal L$ and $T$ is the transpose. 
Explicitly, in terms of the cross-spectral density the OM is expressed as follows
\begin{equation}
O_{lm}(\pmb{\xi})=\frac{1}{4\pi^2}\int\limits_0^{\infty}\int\limits_0^{\infty}e^{il\phi_1}W(\pmb{\xi},\phi_1,\pmb{\xi},\phi_2)e^{-im\phi_2}d\phi_1d\phi_2.
\end{equation}
This implies that the OM can be evaluated from the experimentally measured cross-spectral density \cite{fast}. Alternatively, its individual elements can be measured directly with the help of the Mach-Zehnder interferometer as suggested in  \cite{Roadmap}.

We express  $\lowarrow{O}(\pmb{\xi})= \lowarrow{R}(\pmb{\xi})+ i\lowarrow{I}(\pmb{\xi})$ via its real and imaginary parts:
\begin{align}\label{dec}
&\lowarrow{R}(\pmb{\xi})=\Re [\lowarrow{O}(\pmb{\xi})]= [R_{lm}(\pmb{\xi})],\\&
\lowarrow{I}(\pmb{\xi})=\Im [\lowarrow{O}(\pmb{\xi})] =[I_{lm}(\pmb{\xi})],\label{dec-2}
\end{align} 
yielding symmetric and skew-symmetric matrices, viz., $\lowarrow{R}
(\pmb{\xi})= \lowarrow{R}^{T}(\pmb{\xi})$ and $\lowarrow{I}(\pmb{\xi})= -\lowarrow{I}^{T}(\pmb{\xi})$, with elements $R_{lm}(\pmb{\xi})=|O_{lm}(\pmb{\xi})|\cos[\varphi_{lm}(\pmb{\xi})]$ and $I_{lm}(\pmb{\xi})=|O_{lm}(\pmb{\xi})|\sin[\varphi_{lm}(\pmb{\xi})]$. See Appendix~\ref{appA} for alternative representation of the two parts of the OM. 

To highlight the importance of both parts of the OM we will now show their connection with the OAM flux density associated with the $z$ component of the angular momentum flowing across a surface of constant $z$  \cite{PhysRevA.86.043814}:
\begin{equation}\label{Lzzscalar}
\ell_{zz}(\textbf{r})=\frac{\epsilon_0}{2k} \Im \langle E^*(\textbf{r})\partial_\phi E(\textbf{r})\rangle,
\end{equation}
where $\partial$, $\epsilon_0$ and $k$ denote partial derivative, free-space electric permittivity and wave number, respectively. Indeed, using decomposition (\ref{SpFs}) in Eq.~(\ref{Lzzscalar}) and employing Eq.~(\ref{OM}) yields 
\begin{align}\label{Lzz}
\ell_{zz}(\textbf{r})&= \frac{\epsilon_0}{2k} \Bigl\{\sum\limits_{l\in\mathcal{L}}l R_{ll}(\pmb{\xi})
+ \sum\limits_{l\neq m}^{\infty}l  R_{lm}(\pmb{\xi})\cos[(l-m)\phi] \nonumber\\&
 +\sum\limits_{l\neq m} lI_{lm}(\pmb{\xi})\sin[(l-m)\phi]
 \Bigl\},
\end{align}
where $l,m\in \mathcal{L}$ in the second and third summation. All the elements of the OM contribute to $
\ell_{zz}(\textbf{r})$. The weighed elements $lR_{lm}(\pmb{\xi})$ and $lI_{lm}(\pmb{\xi})$ can be regarded as the coefficients in the Fourier series of $
\ell_{zz}(\textbf{r})$ with respect to angle $\phi$. The first of the three terms superposes the weights of the OAM modes, and, as will be shown in the following sections, the second (third) term represents their in-phase (out-of-phase) correlations, which can also be associated with linear (circular) orbitalization. However, the off-diagonal terms ($l \neq m$) vanish upon integration of the OAM flux density over the transverse plane in obtaining the total OAM flux 
\begin{align}
  L_{zz}
(z) = \iint \ell_{zz}(\mathbf{r})\rho d\rho d\phi=\sum_{l\in\mathcal{L}} L_{zz}^{(l)}(z),
\end{align}
where $L_{zz}^{(l)}(z)=l\pi\epsilon_0\int  R_{ll}(\pmb{\xi}) \rho d\rho/k$ is the OAM flux associated with an individual OAM-mode $E_l(\pmb{\xi})e^{il\phi}$. Thus, only the diagonal elements of the OM contribute to the total OAM flux. This contrasts with the spin angular momentum flux, which is determined exclusively by the off-diagonal of the polarization matrix \cite{PhysRevA.86.043814}. 

For later use, we introduce the Schatten 1- and 2-norms of a matrix $\lowarrow{X}$ \cite{Watrous}:  
\begin{align}\label{Schatten}
    \|\lowarrow{X}\|_1=\text{tr}[(\lowarrow{X}^\dagger \lowarrow{X})^{1/2}], \quad 
    \| \lowarrow{X} \|_2=\bigl[\text{tr}(\lowarrow{X}\lowarrow{X}^\dagger)\bigr]^{1/2},
\end{align} 
where $\text{tr}$ is trace and $\dagger$ stands for Hermitian adjoint. In particular, the 2-norm is equal to the Frobenius norm, while for PSD matrices the 1-norm coincides with trace.

\section{Canonical forms}

\subsection{Real part and beam correlation geometry}

Let us now discuss canonical forms of $\lowarrow{R}(\pmb{\xi})$, $\lowarrow{I}(\pmb{\xi})$ and then $\lowarrow{O}(\pmb{\xi})$. Symmetric and PSD matrix $\lowarrow{R}(\pmb{\xi})$ can be diagonalized to form
\begin{equation}
\lowarrow{M}(\pmb{\xi}) 
=\text{diag} [\mu_1(\pmb{\xi}),\mu_2(\pmb{\xi}),...,\mu_L(\pmb{\xi})],
\end{equation}
where eigenvalues $\mu_n(\pmb{\xi})\geq 0$, with ordering set as $\mu_n(\pmb{\xi})\geq \mu_m(\pmb{\xi})$ for $n\leq m$. The diagonalization is achieved by a real orthogonal matrix $\lowarrow{A}(\pmb{\xi})$:  
\begin{equation}\label{M}
\lowarrow{M}(\pmb{\xi}) 
=\lowarrow{A}^T(\pmb{\xi})
\lowarrow{R}(\pmb{\xi})\lowarrow{A}(\pmb{\xi}),
\end{equation}
whose columns are 
eigenvectors $\vec{a}_n(\pmb{\xi})$ of $\lowarrow{R}(\pmb{\xi})$ 
\begin{equation}\label{EDR}
\lowarrow{R}(\pmb{\xi})\vec{a}_n(\pmb{\xi})=\mu_n(\pmb{\xi})\vec{a}_n(\pmb{\xi}).
\end{equation}
Notice that above $n\in\{1,\dots,L\}\neq\mathcal{L}$, and we have chosen a different index from that used in Eq.~(\ref{SpFs}) to highlight this difference. Since each $L$D vector $\vec{a}_n(\pmb{\xi})$ is real-valued as an eigenvector of a real symmetric matrix, and thus all its components are in phase, it may be regarded as representing a linear orbitalization state, in terminological analogy with a linear state of polarization. Furthermore, the associated OM is given by the outer product $\vec{a}_n(\pmb{\xi})\vec{a}_n^\mathrm{T}(\pmb{\xi})$, which implies that the corresponding state is fully correlated \cite{OErandom}, while the relation $\vec{a}_n^\mathrm{T}(\pmb{\xi})\vec{a}_m(\pmb{\xi})=0$ for $n\neq m$ shows that different states are mutually orthogonal. Multiplying Eq.~(\ref{EDR}) with $\vec{a}_n^\mathrm{T}(\pmb{\xi})$ from the left yields
\begin{align}
    \mu_n(\pmb{\xi})=\langle |\vec{a}_n^\mathrm{T}(\pmb{\xi}) \vec{E}(\pmb{\xi})|^2\rangle. \label{mu}
\end{align}
This shows that the eigenvalue $\mu_n(\pmb{\xi})$ can be interpreted as the mean squared magnitude of the projection of $\vec{E}(\pmb{\xi})$ onto the eigenvector $\vec{a}_n(\pmb{\xi})$, characterizing the in-phase correlation strength associated with the corresponding linear orbitalization state. In addition, Eq.~(\ref{M}) implies that the factorization 
\begin{align}\label{Rdec}
\lowarrow{R}(\pmb{\xi})=\sum\limits_{n=1}^L\mu_n(\pmb{\xi})\vec{a}_n(\pmb{\xi})\vec{a}_n^T(\pmb{\xi})
\end{align}
takes place, showing that $\lowarrow{R}(\pmb{\xi})$ is composed of a set of orthogonal, linearly orbitalized modes whose weights are determined by $\mu_n(\pmb{\xi})$. 

\begin{figure}
    \centering
\includegraphics[width=0.75\columnwidth]{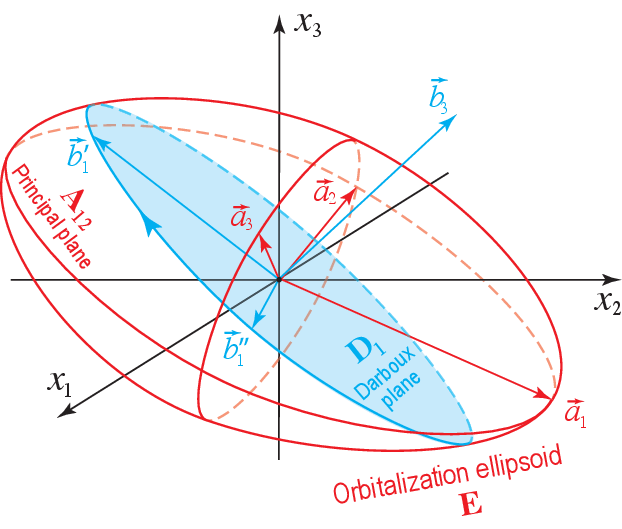}
    \caption{Orbitalization state for $L=3$: ellipsoid $\textbf{E}(\pmb{\xi})$ with the principal plane $\textbf{A}_{12}(\pmb{\xi})$, and the Darboux plane $\textbf{D}_1(\pmb{\xi})$.  
    }
    \label{Fig1} 
\end{figure}

We proceed to show that an ellipsoid can be associated with $\lowarrow{R}(\pmb{\xi})$ in $\mathbb{R}^L$, whose principal axes are aligned with $\vec{a}_n(\pmb{\xi})$. 
Defining 
quadratic forms denoted by $q_{X}(\pmb{\xi},\vec{x})=\vec{x}^T\lowarrow{X}(\pmb{\xi})\vec{x}$ for matrix $\lowarrow{X}(\pmb{\xi})$ with any vector $\vec{x}\in \mathbb{R}^L$, we notice that $q_{O}(\pmb{\xi},\vec{x})=q_R(\pmb{\xi},\vec{x})$ since $q_I(\pmb{\xi},\vec{x})=0$ due to skew-symmetricity. In other words, $\lowarrow{I}(\pmb{\xi})$ does not contribute to the quadratic form of $\lowarrow{O}(\pmb{\xi})$. Then, using eigen-decomposition (\ref{EDR}) together with rotation $\vec{x}'(\pmb{\xi})=\lowarrow{A}^T(\pmb{\xi})\vec{x}$ from the space-fixed to the ellipsoid-fixed frame yields
\begin{align}
&q_{R}(\pmb{\xi},\vec{x})=\vec{x}^T\lowarrow{R}(\pmb{\xi})\vec{x}=
\sum\limits_{n=1}^L \mu_n(\pmb{\xi}) [x^\prime_n(\pmb{\xi})]^2\geq 0, \label{qR} \\&
q_{R^{-1}}(\pmb{\xi},\vec{x})=\vec{x}^T\lowarrow{R}^{-1}(\pmb{\xi})\vec{x}=\sum\limits_{n=1}^L \frac{[x^\prime_n(\pmb{\xi})]^2}{\mu_n(\pmb{\xi})} \geq 0, \label{qR-1}
\end{align}
where $x_n^\prime(\pmb{\xi})$ are the components of $\vec{x}^\prime (\pmb{\xi})$. Thus we see that $\lowarrow{R}(\pmb{\xi})$ and its inverse define dual ellipsoids with the same principal axes but with semi-axes  $1/\sqrt{\mu_n(\pmb{\xi})}$ and $\sqrt{\mu_n(\pmb{\xi})}$, respectively. We will use the latter form  to define \textit{orbitalization  ellipsoid} $\textbf{E}(\pmb{\xi})$:
\begin{equation}\label{OrbEll}
\mathbf{E}(\pmb{\xi}) = \bigl\{ \vec{x} \in \mathbb{R}^L \;\big|\; q_{R^{-1}}(\pmb{\xi},\vec{x}) = 1 \bigr\},
\end{equation}
depicting beam's \textit{correlation geometry} associated with $\lowarrow{R}(\pmb{\xi})$. In defining $\mathbf{E}(\pmb{\xi})$, the units of $\vec{x}(\pmb{\xi})$ are set such that the quadratic form is dimensionless. The $L$ principal axes of $\textbf{E}(\pmb{\xi})$ are tilted to directions $\vec{a}_n(\pmb{\xi})$, pointing out the associated linear orbitalization states, and semi-axes are stretched as $\sqrt{\mu_n(\pmb{\xi})}$, indicating the weights of the states. This construction is analogous to the so-called inertia ellipsoid of 3D polarization states \cite{Dennis2004}. Figure \ref{Fig1} shows $\textbf{E}(\pmb{\xi})$ in $\mathbb{R}^3$ with semi-axes $\sqrt{\mu_n(\pmb{\xi})}$, along directions $\vec{a}_n(\pmb{\xi})$, $n=1,2,3$ respectively. Notice also that $q_{R^{-1}}(\pmb{\xi},\vec{x})\neq q_{O^{-1}}(\pmb{\xi},\vec{x})$, i.e., one cannot directly substitute $\lowarrow{O}$ into Eq.~(\ref{qR-1}) to obtain $\textbf{E}(\pmb{\xi})$. 

For deterministic beams the OM factors as $\lowarrow{O}(\pmb{\xi})=\vec{\mathcal{E}}^*(\pmb{\xi})\vec{\mathcal{E}}^T(\pmb{\xi})$, where $\vec{\mathcal{E}}(\pmb{\xi})=\vec{\mathcal{E}}_r(\pmb{\xi})+i\vec{\mathcal{E}}_i(\pmb{\xi})$ is the spatial part of a monochromatic field realization $\vec{\mathcal{E}}(\pmb{\xi},t)=\vec{\mathcal{E}}(\pmb{\xi})e^{-i\omega t}$ \cite{OErandom}. Hence 
\begin{equation}
\lowarrow{R}(\pmb{\xi})=\vec{\mathcal{E}}_r(\pmb{\xi})\vec{\mathcal{E}}_r^T(\pmb{\xi})+\vec{\mathcal{E}}_i(\pmb{\xi})\vec{\mathcal{E}}_i^T(\pmb{\xi}),
\end{equation}
being a sum of two outer products and thus having at most rank 2. In the rank-2 case, $\vec{\mathcal{E}}_r(\pmb{\xi})$ and $\vec{\mathcal{E}}_i(\pmb{\xi})$ are linearly independent, implying that $\mathbf{E}(\pmb{\xi})$ represents an ellipse:  
\begin{align}
    q_{R^{-1}}(\pmb{\xi},\vec{x})= \frac{[x_1^\prime(\pmb{\xi})]^2}{ \mu_1(\pmb{\xi})}+\frac{[x_2^\prime(\pmb{\xi})]^2}{\mu_2(\pmb{\xi})}=1,
\end{align}
which is in agreement with Eq.~(18) of \cite{OE}. Indeed, it can be shown that the form above coincides with the OE, i.e., the elliptic curve traced by $\Re[\vec{\mathcal{E}}(\pmb{\xi},t)e^{-i\omega t}]$. For the rank-1 case, $\vec{\mathcal{E}}_r(\pmb{\xi})$ and $\vec{\mathcal{E}}_i(\pmb{\xi})$ are linearly dependent, such that $\mathbf{E}(\pmb{\xi})$ reduces to a linear state:
\begin{align}
x_1^\prime(\pmb{\xi}) = \pm \sqrt{\mu_1(\pmb{\xi})}, 
\end{align}
being consistent with Eq.~(24) of \cite{OE}.  

\subsection{Imaginary part and beam correlation topology}

\subsubsection{Darboux form}
An initial insight into the structure of $\lowarrow{I}(\pmb{\xi})$ can be gained by observing that it is a coefficient matrix of a 2-form and can be associated with infinitesimal rotations (\cite{ArnoldCM}, p.~167). Indeed, its diagonal elements vanish, while each off-diagonal element $I_{lm}(\pmb{\xi})$ can be represented as a signed area of a parallelogram formed by vectors with magnitudes $\sqrt{|O_{lm}(\pmb{\xi})|}$ in a plane defined by index pair $l$ and $m$, and separated by the angle $\varphi_{lm}(\pmb{\xi})$. When $L\geq 2$ there are $L(L-1)/2$ independent parallelograms associated with $\lowarrow{I}(\pmb{\xi})$, same as half of the number of off-diagonal matrix elements.

\begin{figure}
    \centering
\includegraphics[width=\linewidth]{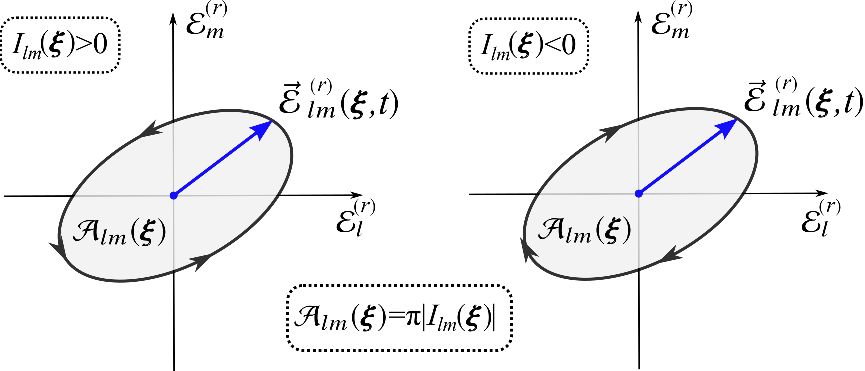}
    \caption{Role of $I_{lm}(\pmb{\xi})$ in the OE traced by the vector $\vec{\mathcal{E}}_{lm}^{(r)}(\pmb{\xi},t)$.}
    \label{fig:ellipses}
\end{figure}

Correspondingly, for a deterministic field, each element $I_{lm}(\pmb{\xi})$ is proportional to the signed area of a 2D orbitalization ellipse \cite{Martinez2024} traced by the tip of the vector $\vec{\mathcal{E}}_{lm}^{(r)}(\pmb{\xi},t)=[\mathcal{E}_l^{(r)}(\pmb{\xi},t),\mathcal{E}_m^{(r)}(\pmb{\xi},t)]$. Here $\mathcal{E}_j^{(r)}(\pmb{\xi},t)=\Re[\mathcal{E}_j(\pmb{\xi})e^{-i\omega t}]$, and $\mathcal{E}_j(\pmb{\xi})$, $j=l,m$, are components of the deterministic OAM-mode vector $\vec{\mathcal{E}}(\pmb{\xi})$. Figure~\ref{fig:ellipses} illustrates this: The area of the ellipse is $\mathcal{A}_{lm}(\pmb{\xi})=\pi|I_{lm}(\pmb{\xi})|$ and $\mathrm{sign}[I_{lm}(\pmb{\xi})]$ determines the traversal sense of the vector $\vec{\mathcal{E}}_{lm}^{(r)}(\pmb{\xi},t)$, with positive (negative) sign corresponding to counterclockwise (clockwise) rotation. Thus, the sign of $I_{lm}(\pmb{\xi})$ is directly connected to the handedness of the generally elliptical 2D orbitalization state, defined here by which component of $\vec{\mathcal{E}}_{lm}^{(r)}(\pmb{\xi},t)$ leads in phase under the adopted $e^{-i\omega t}$ convention: $\mathcal{E}_l^{(r)}(\pmb{\xi},t)$ leads $\mathcal{E}_m^{(r)}(\pmb{\xi},t)$ for $I_{lm}(\pmb{\xi})>0$, and vice versa for $I_{lm}(\pmb{\xi})<0$. In other words, $\lowarrow{I}(\pmb{\xi})$ isolates correlations in $\lowarrow{O}(\pmb{\xi})$ arising from \textit{circulation}. This modal-space circulation should not, however, be confused with the conventional circulation of the electric field around a closed contour in physical space, as appearing in Faraday's law \cite{Jackson}. 

Next, to find the structure of the anti-symmetric  matrix $\lowarrow{I}(\pmb{\xi})$  we transform it to its canonical (Darboux) form (\cite{Horn}, Th. 2.5.13)  
\begin{equation}\label{Dar}
\lowarrow{H}(\pmb{\xi}) 
=\text{diag} [\lowarrow{h_1}(\pmb{\xi}),\lowarrow{h_2}(\pmb{\xi}),...,\lowarrow{h_N}(\pmb{\xi}),\lowarrow{0}_{L-2N}],
\end{equation}
where $N=\text{rank}[\lowarrow{I}(\pmb{\xi})]/2$ and $\lowarrow{0}_{\alpha}$ denotes an $\alpha\times\alpha$ zero block. Each 2$\times$2 block
\begin{equation}\label{DB}
\lowarrow{h_n}(\pmb{\xi})
=\begin{bmatrix}
0 & \eta_n(\pmb{\xi}) \\ -\eta_n (\pmb{\xi})& 0
\end{bmatrix}, \quad n\in \mathcal{N},
\end{equation}
corresponds to a pair $\pm i\eta_n(\pmb{\xi})$, $\eta_n(\pmb{\xi})\in\mathbb{R}$, of nonzero eigenvalues of $\lowarrow{I}(\pmb{\xi})$, with $\mathcal{N}=\{1,...,N\}$. Consequently, the eigenvalues of $\lowarrow{I}(\pmb{\xi})$ are either purely imaginary conjugate pairs, associated with the blocks $\lowarrow{h_n}(\pmb{\xi})$, or zero, linked to the block $\lowarrow{0}_{L-2N}$. Moreover, for odd $L$ at least one eigenvalue is necessarily trivial.

The Darboux form can be obtained by orthogonal transformation (\cite{Horn}, Th. 2.5.17)
\begin{equation}
\lowarrow{H}(\pmb{\xi}) 
=\lowarrow{B}^T(\pmb{\xi})
\lowarrow{I}(\pmb{\xi})\lowarrow{B}(\pmb{\xi}),
\end{equation}
where $\lowarrow{B}(\pmb{\xi})=[\lowarrow{B_1}(\pmb{\xi}),\dots,\lowarrow{B_N}(\pmb{\xi}),\lowarrow{B_0}(\pmb{\xi})]$, with $\lowarrow{B_n}(\pmb{\xi})=[\vec{b}_n'(\pmb{\xi}),\vec{b}_n''(\pmb{\xi})]$, $n\in\mathcal{N}$, and $\vec{b}'_n(\pmb{\xi}),\vec{b}''_n(\pmb{\xi})\in \mathbb{R}^L$ define the real and imaginary parts of the eigenvector $\vec{b}_n(\pmb{\xi})=[\vec{b}'_n(\pmb{\xi})+ i\vec{b}''_n(\pmb{\xi})]/\sqrt{2}$ associated with $+i\eta_n(\pmb{\xi})$. In addition, $\lowarrow{B_0}(\pmb{\xi})=[\vec{b}_{2N+1}(\pmb{\xi}),...,\vec{b}_L(\pmb{\xi})]$ is a $L\times(L-2N)$ matrix whose columns are the real-valued eigenvectors associated with the trivial eigenvalues of $\lowarrow{I}(\pmb{\xi})$. Since $\lowarrow{I}(\pmb{\xi})$ is real, the eigenvector associated with $-i\eta_n(\pmb{\xi})$ is its complex conjugate, $\vec{b}_n^\ast(\pmb{\xi})$. Furthermore, every $\vec{b}_n(\pmb{\xi})$ is also an eigenvector of the Hermitian matrix $i\lowarrow{I}(\pmb{\xi})$, such that eigenvectors corresponding to distinct eigenvalues are orthogonal. Substituting the form of $\vec{b}_n(\pmb{\xi})$ into the eigenvalue equation 
\begin{align}
    \lowarrow{I}(\pmb{\xi})\vec{b}_n(\pmb{\xi})=i\eta_n(\pmb{\xi})\vec{b}_n(\pmb{\xi}), \label{I-eigval}
\end{align}
and separating real and imaginary parts yields
\begin{equation}\label{bilinear}
\lowarrow{I}(\pmb{\xi})\vec{b}'_n(\pmb{\xi})=-\eta _n \vec{b}''_n(\pmb{\xi}), \quad \lowarrow{I}(\pmb{\xi})\vec{b}''_n(\pmb{\xi})= \eta _n \vec{b}'_n(\pmb{\xi}).
\end{equation}
In addition, orthogonality of $\lowarrow{B}(\pmb{\xi})$ 
implies that $\vec{b}_n'(\pmb{\xi})$ and $\vec{b}_n''(\pmb{\xi})$ are orthonormal. 

Consequently, the eigenvector $\vec{b}_n(\pmb{\xi})$ represents an orbitalization state that is a superposition of two mutually orthogonal linear states having equal magnitudes and a relative phase lag of $\pi/2$. This can also be viewed as a circular orbitalization state, in analogy with circular polarization, with the related factoring OM, $\vec{b}_n^\ast(\pmb{\xi})\vec{b}_n^T(\pmb{\xi})$, indicating that the state is fully correlated. Similarly, the conjugate eigenvector $\vec{b}_n^\ast(\pmb{\xi})$ corresponds to a circular state with the opposite handedness. Furthermore, the states described by different eigenvectors are mutually orthogonal in $\mathbb{R}^{2N}$.

Each real-valued vector pair $\vec{b}'_n(\pmb{\xi})$  and $\vec{b}''_n(\pmb{\xi})$ spans a planar subspace of $\lowarrow{I}(\pmb{\xi})$. These Darboux planes are given  as $\textbf{D}_n(\pmb{\xi})=\text{span}\{\vec{b}'_n(\pmb{\xi}),\vec{b}''_n(\pmb{\xi})\}$. Figure \ref{Fig1} shows the only Darboux plane $\textbf{D}_1(\pmb{\xi})$ existing in $\mathbb{R}^3$ spanned by vectors $\vec{b}'_1(\pmb{\xi})$,  $\vec{b}''_1(\pmb{\xi})$ and having normal at the direction of the eigenvector $\vec{b}_3(\pmb{\xi})$ corresponding to the trivial eigenvalue of $\lowarrow{I}(\pmb{\xi})$.
 
Using Eq.~(\ref{bilinear}) together with the components of $\vec{E}(\pmb{\xi})$ along the $n$th Darboux basis vectors, $\zeta_n'(\pmb{\xi})=\vec{b}_n'^T(\pmb{\xi})\vec{E}(\pmb{\xi})$, $\zeta_n''(\pmb{\xi})=\vec{b}_n''^T(\pmb{\xi})\vec{E}(\pmb{\xi})$, we may write 
\begin{align}\label{eta1}
    \eta_n(\pmb{\xi})=\Im[\langle \zeta_n'^\ast(\pmb{\xi})\zeta_n''(\pmb{\xi})\rangle].
\end{align}
Consequently, $\eta_n(\pmb{\xi})$ can be viewed as a measure of out-of-phase correlation between the projections of $\vec{E}(\pmb{\xi})$ onto the linear orbitalization states represented by $\vec{b}_n'(\pmb{\xi})$ and $\vec{b}_n''(\pmb{\xi})$. Employing alternative projections $\zeta_n^+(\pmb{\xi})=\vec{b}_n^T(\pmb{\xi})\vec{E}(\pmb{\xi})$ and $\zeta_n^-(\pmb{\xi})=[\vec{b}_n^\ast(\pmb{\xi})]^T\vec{E}(\pmb{\xi})$ together with Eqs.~(\ref{OM}), (\ref{dec-2}), and (\ref{I-eigval}), the following result can be established:
\begin{align}\label{eta2}
\eta_n(\pmb{\xi})=\frac{1}{2}\left[\langle|\zeta_n^-(\pmb{\xi})|^2\rangle-\langle|\zeta_n^+(\pmb{\xi})|^2\rangle\right].
\end{align}
This relation expresses $\eta_n(\pmb{\xi})$ as the intensity imbalance between the projections of $\vec{E}(\pmb{\xi})$ onto the directions $\vec{b}_n(\pmb{\xi})$ and $\vec{b}_n^\ast(\pmb{\xi})$, which correspond to opposite helicities of circular orbitalization. Both results above are analogous to different expressions of the third Stokes parameter of an   electromagnetic beam \cite{Gil}, which quantifies the degree of circular polarization: Eq.~(\ref{eta1}) is structurally analogous to the representation of this parameter in terms of the Cartesian field components, whereas Eq.~(\ref{eta2}) is analogous to its expression as the intensity difference between the right- and left-hand circularly polarized components. This suggests that each $\eta_n(\pmb{\xi})$ may be regarded as a counterpart of this measure of circular polarization, describing the circulation associated with the $n$th Darboux plane of a structured random beam. In addition, Eq.~(\ref{eta1}) predicts the existence of a 2D polarization-ellipse analogue, traced by the (real part of) deterministic portion of the electric vector $[\zeta_n'(\pmb{\xi}),\zeta_n''(\pmb{\xi})]^T$  
lying in the corresponding Darboux plane, and the total number of such ellipses linked to the beam being $N$. 

Each of the  Darboux planes $\textbf{D}_n(\pmb{\xi})$ intersect with the orbitalization ellipsoid $\textbf{E}(\pmb{\xi})$ resulting in a set of $N$ ellipses. Figure \ref{Fig1} shows such an ellipse for the 3D ellipsoid intersecting the Darboux plane $\textbf{D}_1(\pmb{\xi})$, in blue color. Points residing on 
the $n$th Darboux plane satisfy equation $\vec{x}_n=\lowarrow{B}_n(\pmb{\xi})\vec{s}_n$, $\vec{s}_n\in\mathbb{R}^2$. Substituting 
it into
the ellipsoid form, Eq.~(\ref{OrbEll}), yields $\vec{s}_n^T\lowarrow{B}_n^T(\pmb{\xi}) \lowarrow{R}^{-1}(\pmb{\xi})\lowarrow{B}_n(\pmb{\xi})\vec{s}_n=1$. Letting $\vec{v}^T\vec{v}=1$, $\vec{v}\in\mathbb{R}^2$, represent a unit circle on the Darboux plane, the ellipsoid-plane intersection is found to correspond to the transformation $\vec{s}_n=\lowarrow{T}_n(\pmb{\xi})\vec{v}$, where $\lowarrow{T}_n(\pmb{\xi})=[\lowarrow{B}_n^T(\pmb{\xi}) \lowarrow{R}^{-1}(\pmb{\xi})\lowarrow{B}_n(\pmb{\xi})]^{-1/2}$. 

Darboux form (\ref{Dar}) uniquely defines the \textit{orbitalization circulation state} of an $L$-mode OAM carrying, scalar, random beam as the set of signed scalars 
\begin{equation}
\begin{split}
\textbf{C}(\pmb{\xi})= \{ {\eta}_n(\pmb{\xi}) \in \mathbb{R} \; \big| \;  n\in \mathcal{N}\},
\end{split}
\end{equation}
with magnitudes and signs associated with the strengths and directions of circulation, respectively. 
We also introduce  \textit{net orbitalization circulation} of the beam, 
$C(\pmb{\xi})=\sum\limits_{n\in\mathcal{N}} {\eta_n}(\pmb{\xi})$.   
Contributions from different Darboux blocks to $C(\pmb{\xi})$ may be accumulated, partially suppressed, or vanish. Additionally, quantity $N$ counting the number of non-trivial, independent circulations in the beam can be viewed as the \textit{orbitalization circulation index}. For polarization this number can only be zero or one for both beam-like and general 3D fields, but for orbitalization it can take any integer value from zero up to half of $L$. 

Importantly, in the 
deterministic case $\lowarrow{I}(\pmb{\xi})$ reduces to form
\begin{equation}\label{Idet}
\lowarrow{I}(\pmb{\xi})=\vec{\mathcal{E}}_r(\pmb{\xi})\vec{\mathcal{E}}_i^T(\pmb{\xi})-\vec{\mathcal{E}}_i(\pmb{\xi})\vec{\mathcal{E}}_r^T(\pmb{\xi}),
\end{equation}
and can be shown to have $\mathrm{rank}[\lowarrow{I}(\pmb{\xi})]=2$ when $\vec{\mathcal{E}}_r(\pmb{\xi})$ and $\vec{\mathcal{E}}_i(\pmb{\xi})$ are linearly independent and $\mathrm{rank}[\lowarrow{I}(\pmb{\xi})]=0$ if they align (see Appendix~\ref{appB}).  In the rank-two case, the set $\textbf{C}(\pmb{\xi})$ contains a single element 
and, hence, $C(\pmb{\xi})=\eta_1(\pmb{\xi})$. Then the only Darboux plane merges with that of the OE since they are spanned by the same two vectors, $\vec{\mathcal{E}}_r(\pmb{\xi})$ and $\vec{\mathcal{E}}_i(\pmb{\xi})$, see Eq. (\ref{mu}). In the rank-zero case, the OE degenerates into a linear state and no nontrivial Darboux plane exists.

Thus, the Darboux form of $\lowarrow{I}(\pmb{\xi})$ enables decomposition of total $L$D circulation into a spectrum of $N$ circulations of strengths $|\eta_n(\pmb{\xi})|$ residing in mutually orthogonal planes, specified by vectors $\vec{b}_n(\pmb{\xi})\in \mathbb{C}^L$, $n\in\mathcal{N}$,  
and augmented by $L-2N$ circulation-free modes. Circulation, hence, is inherently pairwise, and remarkably, the particular form of $\lowarrow{I}(\pmb{\xi})$ makes these pairs irreducible and mutually decoupled. Unlike with $\lowarrow{R}(\pmb{\xi})$, whose principal directions specified by vectors $\vec{a}_n(\pmb{\xi})$ decouple all $L$ individual modes, the form of $\lowarrow{I}(\pmb{\xi})$ enforces circulatory coupling of mode pairs, and no such coupling may exist across different Darboux planes. 
Finally, we remark that as long as at least one non-zero Darboux cell exists, the beam has a non-trivial circulatory correlation state. This is analogous to polarization optics, where such structure exists between the Cartesian field components only when a non-trivial circular component (phase lag) is present.

\subsubsection{Darboux forms for $L=2,3,4$.  }

We will now discuss in detail the eigenspace of $\lowarrow{I}(\pmb{\xi})$ for $L=2,3,4$. 
For $L=2$, with two participating OAM-modes in $\vec{E}(\pmb{\xi})=[E_l(\pmb{\xi}),E_m(\pmb{\xi})]^T$, 
the matrix has eigenvalues $\pm i\eta_1(\pmb{\xi})$, with $\eta_1(\pmb{\xi})= I_{lm}(\pmb{\xi})$ describing the strength and handedness of circulation. The corresponding eigenvectors are $\vec{b}_1(\pmb{\xi})=[1,\, i]^T/\sqrt{2}$ and $\vec{b}_1^\ast(\pmb{\xi})$, 
describing 2D circular orbitalization states of opposite handedness, directly analogous to circular polarization states in 2D polarization theory.  

For $L=3$, such that $\vec{E}(\pmb{\xi})=[E_l(\pmb{\xi}),E_m(\pmb{\xi}),E_p(\pmb{\xi})]^T$, the matrix $\lowarrow{I}(\pmb{\xi})$ has again one pair of nonzero eigenvalues, $\pm i\eta_1(\pmb{\xi})$, with
$|\eta_1(\pmb{\xi})|= \| \lowarrow{I}(\pmb{\xi})\|_2/\sqrt{2}$ and $\| \|_2$ representing the Frobenius norm. In addition, $\lowarrow{I}(\pmb{\xi})$ has one trivial eigenvalue corresponding to the real eigenvector $\vec{b}_{3}(\pmb{\xi})=[I_{mp}(\pmb{\xi}), I_{pl}(\pmb{\xi}), I_{lm}(\pmb{\xi})]^T/\eta(\pmb{\xi})$ found by solving $\lowarrow{I}(\pmb{\xi})\vec{b}_{3}(\pmb{\xi})=0$. Therefore, in addition to planar circulation described by $\eta_1(\pmb{\xi})$, the third dimension manifests the axis normal to the unique Darboux plane (see Fig. \ref{Fig1}), spanned by $\vec{b}_3(\pmb{\xi})$. Importantly, 
since $\lowarrow{I}(\pmb{\xi})$ is skew-symmetric, identity
$\lowarrow{I}(\pmb{\xi})\vec{x}=\vec{\omega}(\pmb{\xi})\times\vec{x}$
holds for any vector $\vec{x}\in$ $\mathbb{R}^3$, with $\vec{\omega}(\pmb{\xi})=- \eta_1(\pmb{\xi})\vec{b}_{3}(\pmb{\xi})$. Eigenvectors $\vec{b}_1(\pmb{\xi})=[\vec{b}_1'(\pmb{\xi})+ i \vec{b}_1''(\pmb{\xi})]/\sqrt{2}$ and $\vec{b}_1^\ast(\pmb{\xi})$ 
can be found as follows. Choosing any unit vector $\vec{b}_1'(\pmb{\xi})$ orthogonal to $\vec{b}_3(\pmb{\xi})$, say $\vec{b}_1'(\pmb{\xi})\propto[I_{lp}(\pmb{\xi}), I_{mp}(\pmb{\xi}),0]^T$, and forming $\vec{b}_1''(\pmb{\xi})=\vec{b}_3(\pmb{\xi})\times \vec{b}_1'(\pmb{\xi})$ ensures that $\vec{b}_1'(\pmb{\xi})$, $\vec{b}_1''(\pmb{\xi})$, and $\vec{b}_3(\pmb{\xi})$ are mutually orthogonal, i.e., $\vec{b}_1'(\pmb{\xi})$ and $\vec{b}_1''(\pmb{\xi})$ span the plane orthogonal to $\vec{b}_3(\pmb{\xi})$. Consequently, the eigenvectors associated with the nonzero eigenvalues  $\vec{b}_1(\pmb{\xi})$ and $\vec{b}_1^\ast(\pmb{\xi})$ 
represent circular orbitalization states of opposite handedness in the unique Darboux plane. Notice also that the eigenvector $\vec{b}_3(\pmb{\xi})$ is analogous to the spin angular momentum vector of 3D polarization states \cite{Dennis2004,Gil2019}. In the deterministic case discussed above, the Darboux plane coincides with the OE plane and $\vec{b}_3(\pmb{\xi})$ is therefore normal to it, in direct analogy with the spin vector being normal to the polarization-ellipse plane of a fully polarized 3D field \cite{Dennis2004}. 

For $L=4$, with $\vec{E}(\pmb{\xi})=[E_l(\pmb{\xi}),E_m(\pmb{\xi}),E_p(\pmb{\xi}),E_q(\pmb{\xi})]^T$, two Darboux planes appear, associated with two pairs of purely imaginary conjugate eigenvalues $\pm i \eta_{1}(\pmb{\xi})$ and $\pm i \eta_{2}(\pmb{\xi})$. The corresponding eigenvectors represent two mutually orthogonal pairs of circular orbitalization states in $\mathbb{R}^4$. Here 
\begin{align}\label{4roots}
|\eta_{n}(\pmb{\xi})|&=  \frac{1}{2}\Bigl\{ \| \lowarrow{I}(\pmb{\xi})\|_2^2 \nonumber\\&+(-1)^{n+1} \sqrt{\| \lowarrow{I}(\pmb{\xi})\|_2^4-16\text{Pf}^2[\lowarrow{I}(\pmb{\xi})]}\Bigr\}^{1/2}
\end{align}
for $n=1,2$, and
\begin{align}
\text{Pf}[\lowarrow{I}(\pmb{\xi})]=I_{lm}(\pmb{\xi})I_{pq}(\pmb{\xi})-I_{lp}(\pmb{\xi})I_{mq}(\pmb{\xi}) +I_{lq}(\pmb{\xi})I_{mp}(\pmb{\xi})
\end{align}
is the Pfaffian \cite{Bourbaki1973} of $\lowarrow{I}(\pmb{\xi})$. In addition, the signs of $\eta_{1}(\pmb{\xi})$ and $\eta_{2}(\pmb{\xi})$ specify the handedness of the two circular-orbitalization contributions. If $\text{Pf}[\lowarrow{I}(\pmb{\xi})]=0$, then $|\eta_{1}(\pmb{\xi})|=\| \lowarrow{I}(\pmb{\xi})\|_2/\sqrt{2}$ and $ \eta_{2}(\pmb{\xi})=0$, so that only one nontrivial Darboux plane, and hence one pair of circular orbitalization states, remains. On the other hand, if $\| \lowarrow{I}(\pmb{\xi})\|_2^2=4|\text{Pf}[\lowarrow{I}(\pmb{\xi})]|$, then $|\eta_{1}(\pmb{\xi})|=|\eta_{2}(\pmb{\xi})|=\| \lowarrow{I}(\pmb{\xi})\|_2/2$, corresponding to equal-strength circular orbitalization in the two Darboux planes, i.e., an isoclinic circulation structure. In this case equal signs of $\eta_{1}(\pmb{\xi})$ and $\eta_{2}(\pmb{\xi})$ give $C(\pmb{\xi})=\pm 2|\eta_{1}(\pmb{\xi})|$, whereas opposite signs yield $C(\pmb{\xi})=0$. Importantly, the corresponding deterministic form has rank at most two irrespective of $L$ and must prescribe a single OE in the 4D space. Hence, while the $L=2$ and $L=3$ cases can contain at most one nontrivial Darboux plane also for a random field, $L=4$ is the first case where partial correlation permits two independent Darboux planes and, correspondingly, two pairs of circular orbitalization states. In other words, randomness can cause the total circulation to decompose into two mutually orthogonal circulations whose planes need not coincide with that of the 4D OE. 


\subsubsection{Invariant tori}

We will now consider the behavior of the Darboux form $\lowarrow{H}(\pmb{\xi})$. Let $\mathbb{R}^{2N}$ be equipped with coordinates $x_j$, $j=1,\dots,2N$, grouped into $N$ Darboux planes $(x_{2n-1},x_{2n})$, $n\in\mathcal N$. In each plane, the antisymmetric part of the dynamics is represented by the standard Darboux block $\lowarrow{h_n}(\pmb{\xi})$ as in Eq.~(\ref{DB}). We notice at once that being purely antisymmetric, with zero diagonal entries, $\lowarrow{h_n}(\pmb{\xi})$ does not admit factorization as a single outer product of vectors. Indeed, $
\det[\lowarrow{h_n}(\pmb{\xi})] = \eta_n^2(\pmb{\xi}),
$ so for $\eta_n(\pmb{\xi})\neq 0$ the matrix has rank two. Since any outer product has rank at most one, such a factorization is impossible. Segregated from the real part, the Darboux block defines a rotation on a plane. Indeed, the autonomous linear system generated by $\lowarrow{h_n}(\pmb{\xi})$ has form 
\begin{equation}\label{Ddyn}\begin{cases}\dot{x}_{2n-1}(s)=\eta_n(\pmb{\xi}) x_{2n}(s),  \\ \dot{x}_{2n}(s)=-\eta_n(\pmb{\xi})x_{2n-1}(s),\end{cases} \quad n\in\mathcal N, \end{equation}
where the dot denotes differentiation with respect to the evolution parameter $s$ of the Darboux flow. For fixed $\pmb{\xi}$, $s$ is an auxiliary parameter whose units are inverse to those of $\eta_n(\pmb{\xi})$.
Since 
\begin{equation}
\begin{split}
\frac{d}{ds}[&x_{2n-1}^2(s) + x_{2n}^2(s)]
\\ &= 2x_{2n-1}(s)\dot{x}_{2n-1}(s) + 2x_{2n}(s)\dot{x}_{2n}(s)
= 0,
\end{split}
\end{equation}
this system preserves the quadratic invariant
\begin{equation}\label{DBc}
x_{2n-1}^2(s) + x_{2n}^2(s)=R_n^2(\pmb{\xi})=\mathrm{const}.
\end{equation}
Differentiating once more gives
\begin{equation}
\begin{cases}
\ddot{x}_{2n-1}(s) = -\eta_n^2(\pmb{\xi})\, x_{2n-1}(s), \\
\ddot{x}_{2n}(s) = -\eta_n^2(\pmb{\xi})\, x_{2n}(s),
\end{cases} \quad n\in\mathcal{N}.
\end{equation}
Hence each coordinate satisfies the harmonic oscillator equation and its general solution has form 
\begin{equation}\label{flow}
\begin{cases}
x_{2n-1}(s) = R_n(\pmb{\xi}) \cos[\Psi_n(\pmb{\xi})-\eta_n(\pmb{\xi}) s ], \\
x_{2n}(s) = R_n(\pmb{\xi}) \sin[\Psi_n(\pmb{\xi})-\eta_n(\pmb{\xi}) s ],
\end{cases} 
\end{equation}
where $n\in\mathcal{N}$. Eliminating 
the evolution parameter $s$ explicitly shows that each trajectory lies on a circle of radius $R_n(\pmb{\xi})$ in the corresponding Darboux plane, traced by the tip of the vector
\begin{align}
    \vec{x}_n(s)=R_n(\pmb{\xi})&\big\{\cos[\Psi_n(\pmb{\xi})-\eta_n(\pmb{\xi})s]\vec{b}_n'(\pmb{\xi})\nonumber\\
    &+\sin[\Psi_n(\pmb{\xi})-\eta_n(\pmb{\xi})s]\vec{b}_n''(\pmb{\xi})\big\}.
\end{align}
Although the Darboux flow preserves the radius in each plane, the corresponding skew-symmetric block fixes neither its value nor the initial angular position. To relate these otherwise free constants to the field statistics, we define them as follows. For each field realization, $\zeta_n'(\pmb{\xi})$ and $\zeta_n''(\pmb{\xi})$ are the two components of the orthogonal projection onto the $n$th Darboux plane. We assign the radius using the total mean intensity carried by the projection, as
\begin{align}\label{Rn}
R_n^2(\pmb{\xi})=\langle |\zeta_n'(\pmb{\xi})|^2\rangle +\langle |\zeta_n''(\pmb{\xi})|^2\rangle.
\end{align}
Further, $\Psi_n(\pmb{\xi})$ specifies the initial angular position of $\vec{x}_n(s)$, such that it is the oriented angle from $\vec{b}_n'(\pmb{\xi})$ to $\vec{x}_n(0)$. We may assign this otherwise free initial angle to the phase of the correlation $\langle \zeta_n'^\ast(\pmb{\xi})\zeta_n''(\pmb{\xi})\rangle$:
\begin{align}\label{Psin}
\Psi_n(\pmb{\xi})=\mathrm{arg}[\langle \zeta_n'^\ast(\pmb{\xi})\zeta_n''(\pmb{\xi})\rangle],
\end{align}
such that the correlation phase selects the initial point of the flow on the $n$th circle. 
In addition, the magnitude of quantity $\eta_n(\pmb{\xi})$ determines the angular velocity of the Darboux flow, 
parametrized by $s$, while its sign 
specifies the sense of rotation. 

Since the Darboux planes are mutually orthogonal, combining the dynamics from each of them produces a manifold in 2$N$D, termed here the \textit{orbitalization torus}:
\begin{equation}\label{Torus}
\textbf{T}^N(\pmb{\xi})=\{x\in\mathbb{R}^{2N}: x_{2n-1}^2+x_{2n}^2=R_n^2(\pmb{\xi}),\; n\in \mathcal{N}\},
\end{equation}
or $\textbf{T}^N(\pmb{\xi})= S^1[R_1(\pmb{\xi})]\times\cdots\times S^1[R_N(\pmb{\xi})]$, where $S^1[R_n(\pmb{\xi})]$ is a circle with radius $R_n(\pmb{\xi})$ residing on Darboux plane $\textbf{D}_n(\pmb{\xi})$. Notice that the product remains a torus also in the equal-radius case, $R_n(\pmb{\xi})=R(\pmb{\xi})$ for all $n \in \mathcal{N}$ (see, e.g., p.~138 of \cite{LaValle}). In addition, for $L=2,3$ the imaginary part of the OM has rank at most two, so that $N\leq 1$. When $N=1$, the torus reduces to a single circle, whereas for $N=0$ it vanishes altogether. This also implies that the higher-dimensional toroidal structure is absent in analogous formulations for random 2D and 3D polarization states.

Both the radii $R_n(\pmb{\xi})$ and the frequencies of oscillation $\eta_n(\pmb{\xi})$ are generally different for different $n$. The former determine the geometry of $\textbf{T}^N(\pmb{\xi})$, whereas the winding behavior of the Darboux trajectory depends on the relation among the latter. The curve is closed if $\eta_p(\pmb{\xi})/\eta_q(\pmb{\xi})\in\mathbb{Q}$, i.e., $\eta_p(\pmb{\xi})$ and $\eta_q(\pmb{\xi})$ are commensurate (rationally dependent) frequencies, $p,q\in\mathcal{N}$, whereas incommensurate (rationally independent) frequencies produce dense coverage of the entire torus.

In the particular case of $N=2$, the orbitalization torus is a Clifford-type 2-torus $\textbf{T}^2(\pmb{\xi})$ in $\mathbb{R}^4$, which can be visualized in $\mathbb{R}^3$ by stereographic projection \cite{Hynd2009} (see Appendix~\ref{appC}). 
Figure \ref{InvariantTori} illustrates the dynamics of system (\ref{Ddyn}) for $N=4$ and with $R_2(\pmb{\xi})/R_1(\pmb{\xi})=0.2$.  Figure \ref{InvariantTori}(A) shows the isoclinic case, $\eta_1(\pmb{\xi})=\eta_2(\pmb{\xi})$, with the projected flow trajectory  degenerating to a circle,
known as Villarceau circle \cite{Villarceau1848}. Similar to how a plane cuts through a cone to form a conic section \cite{Coxeter1969}, a pair of Villarceau circles is a toric section formed by a plane passing through the center of the torus and touching it tangentially at two opposite points \cite{Banchoff1978}. 
Further, Fig. \ref{InvariantTori}(B) shows the case of a winding trajectory with $\eta_2(\pmb{\xi})/\eta_1(\pmb{\xi})=10$; Fig. \ref{InvariantTori}(C) illustrates a possible  precession  with $\eta_2(\pmb{\xi})/\eta_1(\pmb{\xi})=1.1$, the curve closing after 10 cycles; and Fig. \ref{InvariantTori}(D) gives the case of incommensurate frequencies,  $\eta_2(\pmb{\xi})/\eta_1(\pmb{\xi})=6\pi$, also exhibiting small precession from cycle to cycle but without closure. In the fully correlated case, circulation only in one plane is possible, hence cases in Figs. \ref{InvariantTori} (B)-(D) are only pertinent to random beams, while (A) reduces to a single-plane circular flow for a deterministic beam with $N=1$. 

A convenient 2D representation of the Darboux flow trajectories for $N\geq 2$ can be obtained by means of a spirograph, in which the projected trajectory is the superposition of circular motions,
\begin{equation}
X(\pmb{\xi},s)+iY(\pmb{\xi},s)=\sum_{n=1}^{N}R_n(\pmb{\xi}) e^{i\eta_n(\pmb{\xi}) s}.
\end{equation}
For commensurate frequencies the resulting planar curve is closed and exhibits  spirographic patterns, whereas incommensurate frequencies produce quasiperiodic trajectories. This construction connects the projected Darboux geometry with the classical theory of roulettes and with Fourier representations of plane curves \cite{Pawley1979}. Figure \ref{Spirograph} presents the spirographic projections of (A)--(D) the same beams as in Fig. \ref{InvariantTori}, and of (F)--(H) some extensions to $L=6$. 


\begin{figure}
    \centering
\includegraphics[width=\columnwidth]{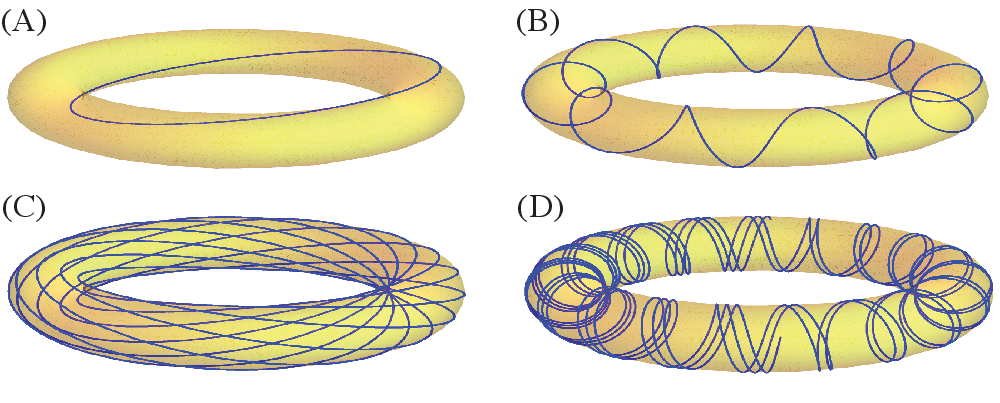}
    \caption{Invariant tori (yellow) and Darboux flow trajectories (blue) for $L=4$, $R_2(\pmb{\xi})/ R_1(\pmb{\xi})=0.2$
    and for $\eta_{2}(\pmb{\xi})/\eta_{1}(\pmb{\xi})$ (A) $1$; (B) $10$; (C) $1.1$; (D) $6\pi$.}
    \label{InvariantTori} 
\end{figure}

\begin{figure*}
  \centering
\includegraphics[width=0.9\textwidth]{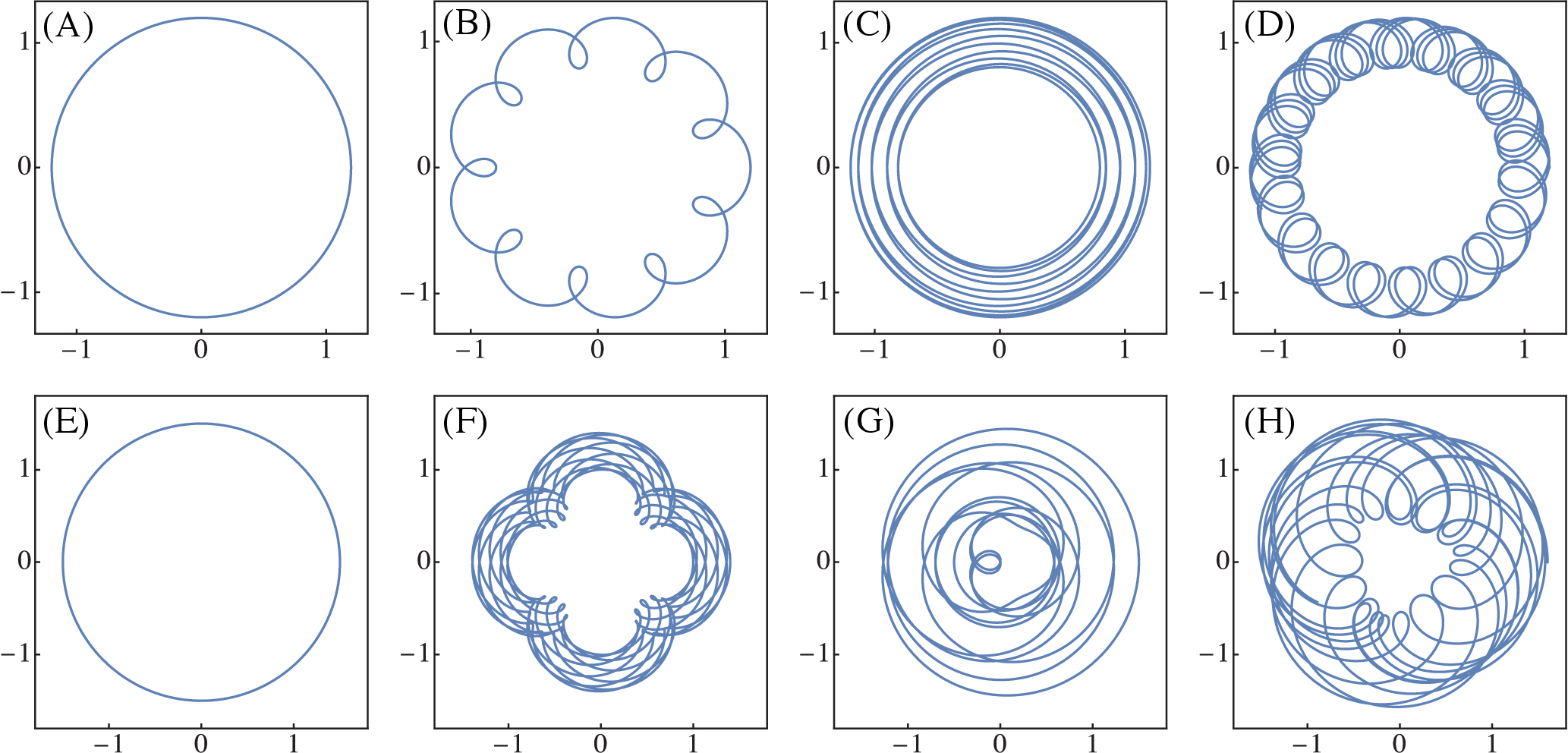}
  \caption{\label{Spirograph} Single-plane (spirograph) projection of Darboux flows for: (A)--(D) the same as in Fig. \ref{InvariantTori}; (E)--(H) $L=6$, with (E) $R_1 = 1$, $R_2 = 0.3$, $R_3 = 0.2$,
$\eta_1 =\eta_2 =\eta_3 = 1$; (F) $R_1 = 1$, $R_2 = R_3 = 0.2$, $\eta_1 = 1$, $\eta_2 = 1.1$, $\eta_3 = 5$; (G) $R_1 = R_2 = R_3 = 0.5$,
$\eta_1 = 1$, $\eta_2 = 1.1$, $\eta_3 = 1.5$; 
(H) $R_1 = 1$, $R_2 = 0.5$, $R_3 = 0.1$,
$\eta_1 = 1$, $\eta_2 = \pi$, $\eta_3 = \text{ln}~10$. All parameters are in arbitrary units. }
\end{figure*}

The Darboux form (\ref{Dar}) and the associated orbitalization torus in Eq.~(\ref{Torus}) characterize the \textit{correlation topology} of the beam. Index $N$ counts the number of active Darboux planes and, within the present Darboux-flow construction, labels the dimension and structure of $\textbf{T}^N(\pmb{\xi})$. Continuous changes of the correlation strengths $\eta_n(\pmb{\xi})$ preserve the correlation topology as long as $N$ remains unchanged; a topological transition occurs only when $N$ changes, corresponding to creation or annihilation of an active Darboux plane. In the construction of $\textbf{T}^N(\pmb{\xi})$, the imaginary part of the OM determines the number of constituent circles $S^1[R_n(\pmb{\xi})]$, and hence the toroidal topology, whereas the full correlation matrix fixes their intensity-weighted radii and thus the geometry of the torus. In this sense, while $\lowarrow{R}(\pmb{\xi})$ defines (continuous) correlation geometry through the deformation of orbitalization ellipsoid $\textbf{E}(\pmb{\xi})$ and of torus $\textbf{T}^N(\pmb{\xi})$, $\lowarrow{I}(\pmb{\xi})$ defines the (discrete) correlation topology through the global structure of torus $\textbf{T}^N(\pmb{\xi})$ with invariant $N$ preserved under continuous deformations and through continuous parameters $\eta_n(\pmb{\xi})$ governing its Darboux flow. 

\subsection{Total orbitalization matrix}

The OM being PSD and Hermitian, can be diagonalized 
as
\begin{equation}
\lowarrow{\Lambda}(\pmb{\xi}) 
=\text{diag} [\lambda_1(\pmb{\xi}),\lambda_2(\pmb{\xi}),...,\lambda_L(\pmb{\xi})]
\end{equation}
via its eigenvalues $\lambda_n(\pmb{\xi})\geq 0$, with ordering $\lambda_p(\pmb{\xi})\geq \lambda_q(\pmb{\xi})$ for $p\leq q$. This is achieved by a unitary matrix $\lowarrow{C}(\pmb{\xi})$, 
\begin{equation}\label{Ospec}
\lowarrow{\Lambda}(\pmb{\xi}) 
=\lowarrow{C}^{\dagger}(\pmb{\xi})
\lowarrow{O}(\pmb{\xi})\lowarrow{C}(\pmb{\xi}),
\end{equation}
whose columns are the complex eigenvectors $\vec{c}_n(\pmb{\xi})$ of $\lowarrow{O}(\pmb{\xi})$, i.e.,  
$
\lowarrow{O}(\pmb{\xi})\vec{c}_n(\pmb{\xi})=\lambda_n(\pmb{\xi})\vec{c}_n(\pmb{\xi}), \; n\in \{1,...,L\}.
$
This leads to the representation
\begin{align}\label{Ospec2}
    \lowarrow{O}(\pmb{\xi})=\sum_{n=1}^L\lowarrow{O}_n(\pmb{\xi}), \quad \lowarrow{O}_n(\pmb{\xi})=\lambda_n(\pmb{\xi})\vec{c}_n(\pmb{\xi})\vec{c}_n^\dagger(\pmb{\xi}),
\end{align}
where each $\lowarrow{O}_n(\pmb{\xi})$ is an $L \times L$, rank-one matrix, representing a fully orbitalized beam and having a corresponding OE, say $\pmb{\epsilon}_n(\pmb{\xi})$ (see also \cite{OErandom}). 

Although decompositions for $\lowarrow{O}(\pmb{\xi})$ and $\lowarrow{R}(\pmb{\xi})$ are similar, the former acts generally in $\mathbb{C}^L$, 
and the latter in $\mathbb{R}^L$ and, therefore, not allowing direct comparison. That is why we have considered the  real and imaginary parts of the OM in $\mathbb{R}^L$ separately and then combined their structures, see Fig. \ref{Fig1}. Equation (\ref{Ospec}) also enables representation of OM via its real embedding (see Appendix~\ref{appD}.) 

It still appears possible to derive general relations involving all members of the three sets. 
Using Eqs.~(\ref{dec}), (\ref{dec-2}), and (\ref{Schatten}), 
one obtains, on omitting $\pmb{\xi}$, expressions:
\begin{equation}\label{PT}
\begin{split}
&\| \lowarrow{O} \|_1= \| \lowarrow{R} \|_1=\sum\limits_{l\in\mathcal{L}}R_{ll}=\sum\limits_{n=1}^L \mu_n, \\ 
&\| \lowarrow{I} \|_1=2\sum_{n=1}^N|\eta_n|, \\&
\| \lowarrow{O} \|_2^2=\| \lowarrow{R}\|^2_2+\| \lowarrow{I}\|^2_2=\sum\limits_{l,m\in\mathcal{L}}|O_{lm}|^2 =\sum\limits_{n=1}^L\lambda_n^2, \\&
\| \lowarrow{R} \|_2^2=\sum\limits_{l\in \mathcal{L}}R_{ll}^2+2\sum_{\substack{l,m\in\mathcal{L}\\ l< m}}|R_{lm}|^2=\sum\limits_{n=1}^L\mu_n^2, \\&
\| \lowarrow{I} \|_2^2 =\sum_{\substack{l,m\in\mathcal{L}\\ l\neq m}}|I_{lm}|^2=2\sum\limits_{n=1}^N\eta_n^2.
\end{split}
\end{equation}
Note that for $\lowarrow{O}(\pmb{\xi})$ and $\lowarrow{R}(\pmb{\xi})$ Schatten norms coincide with traces, i.e., $\| \lowarrow{O}(\pmb{\xi}) \|_1=\text{tr}[\lowarrow{O}(\pmb{\xi})]$ and $\| \lowarrow{R} (\pmb{\xi})\|_1=\text{tr}[\lowarrow{R}(\pmb{\xi})]$. While average radial intensity  $\| \lowarrow{O}(\pmb{\xi}) \|_1$ is determined by $\lowarrow{R}(\pmb{\xi})$, the radial correlation strength $\| \lowarrow{O}(\pmb{\xi}) \|_2$ also depends on  $\lowarrow{I}(\pmb{\xi})$. Equation (\ref{PT}) implies that the total correlation strength is the sum of those arising from the in-phase (alignment) and out-of-phase (circulation) portions. Ratios
\begin{equation}
s_R(\pmb{\xi})=\frac{\| \lowarrow{R}(\pmb{\xi})\|_2}{\| \lowarrow{O}(\pmb{\xi})\|_2}, \quad s_I(\pmb{\xi})=\frac{\| \lowarrow{I}(\pmb{\xi})\|_2}{\| \lowarrow{O}(\pmb{\xi})\|_2},
\end{equation}
give relative contributions of $\lowarrow{R}(\pmb{\xi})$ and $\lowarrow{I}(\pmb{\xi})$ to the 2-norm of $\lowarrow{O}(\pmb{\xi})$, with $s_R^2(\pmb{\xi})+s_I^2(\pmb{\xi})=1$. Then, since 
$\text{tr}[\lowarrow{O}^2(\pmb{\xi})]= \text{tr}[\lowarrow{R}^{2}(\pmb{\xi})]-\text{tr}[\lowarrow{I}^{2}(\pmb{\xi})]$,
the eigenvalues are also linked:
\begin{equation}\label{eigen}
\sum \limits_{n=1}^L \lambda_n^2(\pmb{\xi})=\sum \limits_{n=1}^L \mu_n^2(\pmb{\xi})+2\sum \limits_{n=1}^N \eta_n^2(\pmb{\xi}).
\end{equation}

\section{Degrees of linear and circular orbital anisotropy and stability}

In this section, the counterparts of the degrees of linear and circular polarization and directionality of a vectorial light field 
are derived for a scalar beam in the OAM modal space, using 
canonical forms of $\lowarrow{R}(\pmb{\xi})$, $\lowarrow{I}(\pmb{\xi})$ 
and $\lowarrow{O}(\pmb{\xi})$ introduced in Sec. 3. 

The eigenvalue distribution of $\lowarrow{O}(\pmb{\xi})$ can be quantified by the Degree of Orbital Anisotropy (DOA) \cite{HBT} (see also \cite{Laatikainen:26}) 
\begin{align}\label{Q}
    Q(\pmb{\xi})&=\sqrt{\frac{L}{L-1}\left\{\frac{\|\lowarrow{O}(\pmb{\xi})\|_2^2}{\|\lowarrow{O}(\pmb{\xi})\|_1^2}-\frac{1}{L}\right\}}\nonumber\\
    &=\sqrt{\frac{L}{L-1}\left\{\frac{\sum_{n=1}^L \lambda_n^2(\pmb{\xi})}{[\sum_{n=1}^L \lambda_n(\pmb{\xi})]^2}-\frac{1}{L}\right\}},
\end{align}
which is bounded as $0\leq Q(\pmb{\xi})\leq 1$. The upper limit $Q(\pmb{\xi})=1$ is reached when $\lambda_n(\pmb{\xi})=0$ for $n\geq 2$, i.e., the OM consists of a single $\lowarrow{O}_n(\pmb{\xi})$ in Eq.~(\ref{Ospec2}) and represents a fully orbitalized state. On the other hand, the lower bound $Q(\pmb{\xi})=0$ corresponds to the isotropic case $\lambda_1(\pmb{\xi})=\lambda_n(\pmb{\xi})$, $n\geq 2$, such that all $\lowarrow{O}_n(\pmb{\xi})$ have the same weight, representing a completely isotropic state. The DOA has direct counterparts in polarization theory: For $L=2$ and $L=3$, it is formally identical to the conventional degree of polarization and degree of polarimetric purity, respectively \cite{Gil,Setala2002,Gil2018}.

\subsection{Degrees of linear orbital anisotropy and stability}

\subsubsection{Special cases $L=2,3$} 


The canonical form of $\lowarrow{R}(\pmb{\xi})$ can be employed to introduce the Degree of Linear Orbital Anisotropy (DLOA). Recall Eq.~(\ref{Rdec}), where every real-valued eigenvector $\vec{a}_n(\pmb{\xi})$ represents a linear orbitalization state. For $L=3$, 
the DLOA can be defined in analogy with the degree of linear polarization in 3D \cite{Gil}:
\begin{align}\label{Qlin}
    Q_\mathrm{lin}(\pmb{\xi})=\frac{\mu_1(\pmb{\xi})-\mu_2(\pmb{\xi})}{\sum_{n=1}^3 \lambda_n(\pmb{\xi})}.
\end{align}
Note that because of the identities given in the first line of Eq.~(\ref{PT}), the denominator in Eq.~(\ref{Qlin}) is equal to $\sum_{n=1}^3 \mu_n(\pmb{\xi})$. Similar formulation is valid also for $L=2$, in which case the third eigenvalue is absent from the normalization factor in Eq.~(\ref{Qlin}). Here $\mu_1(\pmb{\xi})-\mu_2(\pmb{\xi})$ measures the imbalance between the two mutually orthogonal linear orbitalization states spanning the principal 2D subspace of $\lowarrow{R}(\pmb{\xi})$, which we will denote by $\textbf{A}_{12}$ (see Fig. \ref{Fig1}). Equal weights, $\mu_1(\pmb{\xi})=\mu_2(\pmb{\xi})$, imply the absence of a dominating linear orbitalization direction within this space, whereas the case $\mu_1(\pmb{\xi})=\lambda_1(\pmb{\xi})$ and $\mu_n(\pmb{\xi})=\lambda_n(\pmb{\xi})=0$, $n=2,3$, corresponds to a fully linear orbitalization state. This is directly analogous to the 3D and 2D polarization constructions, in which the degree of linear polarization is obtained from the difference of the two leading eigenvalues of the real part of the polarization matrix, while the third eigenvalue specifies the degree of directionality, i.e., stability of the plane of the polarization ellipse \cite{Gil2016}. A similar stability measure can be formally defined for $L=3$ as the Degree of Orbital Stability (DOS) by expression 
\begin{align}\label{Qdir}
    Q_\mathrm{stab}(\pmb{\xi})=1-\frac{3\mu_3(\pmb{\xi})}{\sum_{n=1}^3 \lambda_n}(\pmb{\xi}).
\end{align}
The geometric meaning of $Q_\mathrm{stab}(\pmb{\xi})$ can be clarified by considering the spectral decomposition of the OM, Eq.~(\ref{Ospec2}). Substituting 
\begin{align}\label{cn}
\vec{c}_n(\pmb{\xi})=\vec{c}_n'(\pmb{\xi})+i\vec{c}_n''(\pmb{\xi}), 
\end{align}
with $\vec{c}_n'(\pmb{\xi}),\vec{c}_n''(\pmb{\xi})\in\mathbb{R}^L$, to Eq.~(\ref{Ospec2}) and taking the real part gives, for $L=3$,
\begin{align}
    \lowarrow{R}(\pmb{\xi})=\sum_{n=1}^3 \lambda_n(\pmb{\xi})[\vec{c}_n'(\pmb{\xi})\vec{c}_n'^T(\pmb{\xi})+\vec{c}_n''\vec{c}_n''^T].
\end{align}
In particular, we find by using Eq.~(\ref{EDR}) that
\begin{align}\label{mu3}
    \mu_3(\pmb{\xi})=\sum_{n=1}^3 \lambda_n(\pmb{\xi})\left\{[\vec{a}_3^T(\pmb{\xi})\vec{c}_n']^2+[\vec{a}_3^T(\pmb{\xi})\vec{c}_n'']^2\right\}.
\end{align}
On the right hand side above, each term shows the $\lambda_n$-weighted magnitude of the projection of $\vec{c}_n'(\pmb{\xi})$ and $\vec{c}_n''(\pmb{\xi})$ to the direction of $\vec{a}_3(\pmb{\xi})$. Noting that $\vec{a}_3(\pmb{\xi})\perp\textbf{A}_{12}$, the form above shows that $\mu_3(\pmb{\xi})$ represents the total contribution of all $\lowarrow{O}_n(\pmb{\xi})$ outside of $\textbf{A}_{12}$, i.e., its complement $\textbf{A}_{12}^{\perp}$. The corresponding ellipse $\pmb{\epsilon}_n(\pmb{\xi})$ is instead contained in the plane spanned by $\vec{c}_n'(\pmb{\xi})$ and $\vec{c}_n''(\pmb{\xi})$, which is in general different from $\textbf{A}_{12}$ that formed by $\vec{a}_1(\pmb{\xi})$ and $\vec{a}_2(\pmb{\xi})$. However, since every term on the right-hand side is non-negative, the case $\mu_3(\pmb{\xi})=0$ implies that $\vec{c}_n'(\pmb{\xi})$ and $\vec{c}_n''(\pmb{\xi})$ lie in $\mathrm{span}[\vec{a}_1(\pmb{\xi}),\vec{a}_2(\pmb{\xi})]$ for all $n=1,2,3$. In this case, all $\pmb{\epsilon}_n(\pmb{\xi})$ are confined to $\textbf{A}_{12}$, which is also the plane specified by the two largest semi-axes of $\textbf{E}(\pmb{\xi})$. 
Consequently, for $L=3$ the measure $Q_\mathrm{stab}(\pmb{\xi})$ quantifies the degree of common-plane alignment of these $\pmb{\epsilon}_n(\pmb{\xi})$. The maximum $Q_\mathrm{stab}(\pmb{\xi})=1$ is obtained when $\mu_3(\pmb{\xi})=0$ and corresponds to complete alignment to a common plane, whereas $Q_\mathrm{stab}(\pmb{\xi})=0$ requires that $\mu_1(\pmb{\xi})=\mu_2(\pmb{\xi})=\mu_3(\pmb{\xi})$ and no principal 2D subspace is distinguished. This is analogous to the degree of directionality in 3D polarization theory \cite{Gil2016}, with the distinction that the modal subspace considered here has no direct connection to the physical propagation direction of the beam. 


\subsubsection{Case $L>3$}
To extend the quantities above to the general case $L>3$, we start by noting that $\textbf{A}_{12}$ remains the same. However, instead of a single normal direction, $\textbf{A}_{12}^{\perp}$ now has dimension $L-2$, spanned by vectors $\vec{a}_n(\pmb{\xi})$, $n=3,...,L$. Similarly to Eq.~(\ref{mu3}), each eigenvalue $\mu_n(\pmb{\xi})$ can now be expressed as 
\begin{align}\label{mun}
\mu_n(\pmb{\xi})=\sum_{j=1}^L \lambda_j(\pmb{\xi})\left\{[\vec{a}_n^T(\pmb{\xi})\vec{c}_j']^2+[\vec{a}_n^T(\pmb{\xi})\vec{c}_j'']^2\right\},
\end{align}
giving the $\lambda_j$-weighted total strength of the projections of $\vec{c}_j'(\pmb{\xi})$ and $\vec{c}_j''(\pmb{\xi})$ onto the direction of $\vec{a}_n^T(\pmb{\xi})$. Then, the total departure of $\vec{c}_j'(\pmb{\xi})$ and $\vec{c}_j''(\pmb{\xi})$ from $\textbf{A}_{12}$ is given by 
\begin{align}
\Delta(\pmb{\xi})=\sum_{n=3}^L \mu_n(\pmb{\xi}).
\end{align}
Consequently, $\Delta(\pmb{\xi})=0$ when $\mu_n(\pmb{\xi})=0$ for $n=3,...,L$, i.e., only if all $\pmb{\epsilon}_n(\pmb{\xi})$ can be confined to a common 2D subspace. On the other hand, the maximum
\begin{align}\label{maxD}
\mathrm{max}[\Delta(\pmb{\xi})]=\frac{L-2}{L}\sum_{n=1}^L \mu_n(\pmb{\xi})=\frac{L-2}{L}\sum_{n=1}^L \lambda_n(\pmb{\xi})
\end{align}
is attained for a completely isotropic eigenvalue spectrum of $\lowarrow{R}(\pmb{\xi})$, i.e., when $\mu_1(\pmb{\xi})=\mu_n(\pmb{\xi})$ for all $n\geq 2$. Normalizing $\Delta(\pmb{\xi})$ by its maximum value, we may define the generalized $Q_\mathrm{stab}(\pmb{\xi})$ as the complement
\begin{align}\label{QdirL}
    Q_\mathrm{stab}(\pmb{\xi})=1-\frac{\Delta(\pmb{\xi})}{\mathrm{max}[\Delta(\pmb{\xi})]}=1-\frac{L}{L-2}\frac{\sum_{n=3}^L \mu_n(\pmb{\xi})}{\sum_{n=1}^L \lambda_n(\pmb{\xi})}.
\end{align}
Alternatively, denoting the mean weights across states $n = 1,2$ and $n\geq 3$ as
\begin{align}\label{mean}
    \overline{\mu}_{12}(\pmb{\xi})=\frac{\mu_1(\pmb{\xi})+\mu_2(\pmb{\xi})}{2}, \quad \overline{\mu}_\Delta(\pmb{\xi})=\frac{\Delta(\pmb{\xi})}{L-2},
\end{align}
the DOS can be expressed in the form
\begin{align}\label{Qdir2}
    Q_\mathrm{stab}(\pmb{\xi})=\frac{2[\overline{\mu}_{12}(\pmb{\xi})-\overline{\mu}_\Delta(\pmb{\xi})]}{\sum_{n=1}^L\mu_n(\pmb{\xi})}.
\end{align}
Here $Q_\mathrm{stab}(\pmb{\xi})$ is bounded as $0\leq Q_\mathrm{stab}(\pmb{\xi})\leq 1$. The upper limit corresponds to all the $L$ $\pmb{\epsilon}_n(\pmb{\xi})$ residing in $\textbf{A}_{12}$, while at the lower limit no principal subspace can be distinguished. For $L=3$, Eq.~(\ref{QdirL}) reduces to Eq.~(\ref{Qdir}). For $L=2$, the entire OAM space is intrinsically two-dimensional, so that no independent degree of stability is required. Equivalently, $Q_\mathrm{stab}(\pmb{\xi})=1$ may be assigned by convention. This is analogous to 2D polarization, where the polarization structure is necessarily planar and the DOS is therefore trivially maximal.

The extension of $Q_\mathrm{lin}(\pmb{\xi})$ to $L>3$ requires additional consideration, since now vectors $\vec{a}_n(\pmb{\xi})$, $n\geq 3$, represent mutually orthogonal linear orbitalization states and may carry unequal weights. Their total weight is contained in $\Delta(\pmb{\xi})$ and is accounted for by $Q_\mathrm{stab}(\pmb{\xi})$. However, unequal distribution of that weight among different states creates additional linear anisotropy besides that in $\textbf{A}_{12}$. 
Using Eq.~(\ref{mean}) and rearranging the terms in Eq.~(\ref{Rdec}), we may express $\lowarrow{R}(\pmb{\xi})=\lowarrow{R}_{\text{lin}}(\pmb{\xi})+\lowarrow{R}_{\text{iso}}(\pmb{\xi})$, where
\begin{align}\label{Rlin}
\lowarrow{R}_{\text{lin}}(\pmb{\xi})&=
\frac{1}{2}[\mu_1(\pmb{\xi})-\mu_2(\pmb{\xi})][\vec{a}_1(\pmb{\xi})\vec{a}_1^T(\pmb{\xi})-\vec{a}_2(\pmb{\xi})\vec{a}_2^T(\pmb{\xi})] \nonumber\\&
+\sum_{n=3}^L[\mu_n(\pmb{\xi})-\overline{\mu}_\Delta (\pmb{\xi})] \vec{a}_n(\pmb{\xi})\vec{a}_n^T(\pmb{\xi}), \\
\lowarrow{R}_{\text{iso}}(\pmb{\xi})&=
\overline{\mu}_{12}(\pmb{\xi})
\sum_{n=1}^2 \vec{a}_n(\pmb{\xi})\vec{a}_n^T(\pmb{\xi})
+\overline{\mu}_\Delta(\pmb{\xi})\sum_{n=3}^L \vec{a}_n(\pmb{\xi})\vec{a}_n^T(\pmb{\xi}). 
\label{Riso}
\end{align}

Above, the weight of the first term in Eq.~(\ref{Rlin}) is accounted for in Eq.~(\ref{Qlin}) for $L=3$, and it represents linear anisotropy within  $\textbf{A}_{12}$, being maximal when $\mu_2(\pmb{\xi})=0$ and vanishing for $\mu_1(\pmb{\xi})=\mu_2(\pmb{\xi})$. Then, the second term in Eq.~(\ref{Rlin}) represents additional anisotropy carried by the linear orbitalization states in $\textbf{A}_{12}^{\perp}$, vanishing only under complete isotropy in $\textbf{A}_{12}^{\perp}$, i.e., when $\mu_3(\pmb{\xi})=\mu_n(\pmb{\xi})$, $n\geq 4$. The first and second terms in \eqref{Riso} are isotropic within $\textbf{A}_{12}$ and $\textbf{A}_{12}^{\perp}$, respectively, and therefore do not select a preferred linear orbitalization direction within either subspace. 

Consequently, $\lowarrow{R}_{\text{lin}}(\pmb{\xi})$ given by  \eqref{Rlin} can be identified with the component of $\lowarrow{R}(\pmb{\xi})$ associated with linear anisotropy, having the 2-norm
\begin{align}
    \|\lowarrow{R}_\mathrm{lin}(\pmb{\xi})\|_2^2=\frac{1}{2}[\mu_1(\pmb{\xi})-\mu_2(\pmb{\xi})]^2+\sum_{n=3}^L[\mu_n(\pmb{\xi})-\overline{\mu}_\Delta (\pmb{\xi})]^2.
\end{align}
We see that, for fixed $\mathrm{tr}[\lowarrow{R}(\pmb{\xi})]$, the limit $\mathrm{max}\|\lowarrow{R}_\mathrm{lin}(\pmb{\xi})\|_2=\| \lowarrow{R}(\pmb{\xi})\|_1/\sqrt{2}=\| \lowarrow{O}(\pmb{\xi})\|_1/\sqrt{2}$ is reached when $\mu_n(\pmb{\xi})=0$ for $n\geq 2$, i.e., for a completely linearly orbitalized state, while $\mathrm{min}\|\lowarrow{R}_\mathrm{lin}(\pmb{\xi})\|_2^2=0$ is obtained when $\mu_1(\pmb{\xi})=\mu_2(\pmb{\xi})$ and $\mu_3(\pmb{\xi})=\mu_n(\pmb{\xi})$, $n\geq 4$, corresponding to isotropy within both $\textbf{A}_{12}$ and $\textbf{A}_{12}^{\perp}$. Complete isotropy of $\lowarrow{R}(\pmb{\xi})$ constitutes the particular case in which these two common eigenvalues also coincide. This suggests defining the DLOA for $L>3$ as
\begin{equation}\label{QlinL}
\begin{split}
    &Q_\mathrm{lin}(\pmb{\xi})
    =\frac{\sqrt{2}\|\lowarrow{R}_\mathrm{lin}(\pmb{\xi})\|_2}{\|\lowarrow{O}(\pmb{\xi})\|_1} \\&
    =\frac{\big\{[\mu_1(\pmb{\xi})-\mu_2(\pmb{\xi})]^2 +2\sum_{n=3}^L[\mu_n(\pmb{\xi})-\overline{\mu}_\Delta (\pmb{\xi})]^2\big\}^{1/2}}{\sum_{n=1}^L \lambda_n(\pmb{\xi})},
    \end{split}
\end{equation}
which is again bounded as $0\leq Q_\mathrm{lin}(\pmb{\xi})\leq 1$. The first term under the square root measures the imbalance between the two principal linear orbitalization states, whereas the second accounts for anisotropy among the remaining linear states. For a fully linear orbitalization state $Q_\mathrm{lin}(\pmb{\xi})=1$, 
whereas $Q_\mathrm{lin}(\pmb{\xi})=0$ when the eigenvalues are equal separately within $\textbf{A}_{12}$ and $\textbf{A}_{12}^{\perp}$, corresponding to complete isotropy withing both subspaces. For $L=3$ the second term vanishes because $\overline{\mu}_\Delta(\pmb{\xi})=\mu_3(\pmb{\xi})$ and $Q_\mathrm{lin}(\pmb{\xi})$ reduces to Eq.~(\ref{Qlin}). In form, $Q_\mathrm{lin}(\pmb{\xi})$ is the extension of the degree of linear polarization \cite{Gil,Gil2016} and reduces to an identical expression for $L=2,3$.  

\subsection{Degree of circular orbital anisotropy}

\subsubsection{Special cases $L=2,3$} 


The Degree of Circular Orbital Anisotropy (DCOA) can be introduced in analogy with the degree of circular polarization. For both $2D$ and $3D$ polarization states, circular polarization is associated with the imaginary, antisymmetric part of the polarization matrix, and its degree is given by the corresponding circular-polarization strength normalized by the total intensity. Similarly, the circular orbitalization information of the beam is included in the antisymmetric matrix $\lowarrow{I}(\pmb{\xi})$, whose Darboux form can be used to define the DCOA. For $L=3$, the matrix has a single pair of nonzero eigenvalues, $\pm i\eta_1(\pmb{\xi})$, such that $|\eta_1(\pmb{\xi})|$ completely characterizes the strength of circular orbitalization. We may therefore define the DCOA as
\begin{align}\label{Qcirc}
    Q_\mathrm{circ}(\pmb{\xi})=\frac{2|\eta_1(\pmb{\xi})|}{\sum_{n=1}^3 \lambda_n(\pmb{\xi})}.
\end{align}
Here $0\leq Q_\mathrm{circ}(\pmb{\xi})\leq 1$, with the following limitting cases. For a completely circular orbitalization state, which is formed by two orthogonal linear orbitalization states of equal amplitude and relative phase lag $\pi/2$, OM is of the form $\lowarrow{O}(\pmb{\xi})=\|\lowarrow{O}(\pmb{\xi})\|_1\vec{b}_1^\ast(\pmb{\xi})\vec{b}_1^T(\pmb{\xi})$, such that $|\eta_1(\pmb{\xi})|=\|\lowarrow{O}(\pmb{\xi})\|_1/2$ and $Q_\mathrm{circ}(\pmb{\xi})=1$. 
On the other hand, complete lack of circular anisotropy yields $\eta_1(\pmb{\xi})=0$ and thus $Q_\mathrm{circ}(\pmb{\xi})=0$.

\subsubsection{Case $L>3$}
For $L=2,3$ there is, at most, one independent Darboux plane, such that a single scalar $\eta_1(\pmb{\xi})$ specifies the magnitude and direction of circulation, making the definition in Eq.~(\ref{Qcirc}) unambiguous. However, for $L>3$ multiple mutually orthogonal Darboux planes may exist, characterized by $\mathbf{C}(\pmb{\xi})$. The extension of Eq.~(\ref{Qcirc}) to $L>3$ therefore requires specification of how the independent circulation states are to be accounted for. 

As shown in Appendix~\ref{appE}, the 2-norm of $\lowarrow{I}(\pmb{\xi})$ appears as the suitable measure for quantifying circular anisotropy of the beam, as its limiting values have the expected physical interpretations for such a measure: Its lower bound corresponds to complete absence of circular anisotropy, whereas the upper bound yields complete circular anisotropy. Therefore, we define the DCOA as
\begin{align}\label{QcircL}
    Q_\mathrm{circ}(\pmb{\xi})=&\frac{\sqrt{2}\|\lowarrow{I}(\pmb{\xi})\|_2}{\|\lowarrow{O}(\pmb{\xi})\|_1}=\frac{2\left[\sum_{n=1}^N \eta_n^2(\pmb{\xi})\right]^{1/2}}{\sum_{n=1}^L \lambda_n(\pmb{\xi})}.
\end{align}
Here $0\leq Q_\mathrm{circ}(\pmb{\xi})\leq 1$, with $Q_\mathrm{circ}(\pmb{\xi})=0$ implying complete lack of circular anisotropy and $Q_\mathrm{circ}(\pmb{\xi})=1$ indicating full circular anisotropy, i.e., a factoring fully circular orbitalization state. In addition, for $L=2,3$ the above form reduces to Eq.~(\ref{Qcirc}).

\subsection{Relation with degree of orbital anisotropy}

The three degrees introduced above can be related to the DOA $Q(\pmb{\xi})$. 
We start from Eq.~(\ref{eigen}) and decompose the quadratic contribution of the eigenvalues of $\lowarrow{R}(\pmb{\xi})$ using a two-group ($\{\mu_1,\mu_2\}$ and $\{\mu_3,...,\mu_L\}$) K\"onig-Huygens theorem \cite{Scheffe1959}:
\begin{align}\label{Eq105}
    &\sum_{n=1}^L \mu_n^2(\pmb{\xi})=\frac{1}{L}\left[\sum_{n=1}^L \mu_n(\pmb{\xi})\right]^2+\frac{1}{2}[\mu_1(\pmb{\xi})-\mu_2(\pmb{\xi})]^2\nonumber\\
    &+\sum_{n=3}^L[\mu_n(\pmb{\xi})-\overline{\mu}_\Delta(\pmb{\xi})]^2+\frac{2(L-2)}{L}[\overline{\mu}_{12}(\pmb{\xi})-\overline{\mu}_\Delta(\pmb{\xi})]^2.
\end{align}
Formula above gives decomposition of the sum of squared eigenmode weights of $\lowarrow{R}(\pmb{\xi})$ split into four terms (from left to right): the contribution associated with the average weight among all the modes, the internal split of the dominating mode pair; the internal spread of the rest of the modes; and separation between the two groups' centers.

Using Eqs.~(\ref{Qdir2}) and (\ref{QlinL}), we obtain from Eq.~(\ref{Eq105}) that
\begin{align}
    \frac{\sum_{n=1}^L\mu_n^2(\pmb{\xi})}{[\sum_{n=1}^L\mu_n(\pmb{\xi})]^2}=\frac{1}{L}+\frac{1}{2}Q_\mathrm{lin}^2(\pmb{\xi})+\frac{L-2}{2L}Q_\mathrm{stab}^2(\pmb{\xi}).
\end{align}
Note that due to form of \eqref{Qdir2} $Q_{\text{dir}}(\pmb{\xi})$ can be interpreted as degree of separation between the mean weights of the two groups: the dominating pair of OAM modes and the rest of the them.
This expression, together with the definition of $Q_\mathrm{circ}(\pmb{\xi})$ in Eq.~(\ref{QcircL}) and Eq.~(\ref{eigen}) yield
\begin{align}
    \frac{\sum_{n=1}^L\lambda_n^2(\pmb{\xi})}{[\sum_{n=1}^L \lambda_n(\pmb{\xi})]^2}&=\frac{1}{L}+\frac{1}{2}Q_\mathrm{lin}^2(\pmb{\xi})+\frac{1}{2}Q_\mathrm{circ}^2(\pmb{\xi})\nonumber\\&+\frac{L-2}{2L}Q_\mathrm{stab}^2(\pmb{\xi}).
\end{align}
Substituting the expression above into the definition of $Q(\pmb{\xi})$, Eq.~(\ref{Q}), gives
\begin{align}\label{Q-relation}
    Q^2(\pmb{\xi})=\frac{L}{2(L-1)}Q_\mathrm{ell}^2(\pmb{\xi})+\frac{L-2}{2(L-1)}Q_\mathrm{stab}^2(\pmb{\xi}),
\end{align}
where 
\begin{equation}
Q_\mathrm{ell}(\pmb{\xi})=[Q_\mathrm{lin}^2(\pmb{\xi})+Q_\mathrm{circ}^2(\pmb{\xi})]^{1/2}
\end{equation}
is the Degree of Elliptical Orbital Anisotropy (DEOA), summarizing the linear and circular contributions to the DOA. Thus, the DOA separates into contributions associated with linear anisotropy, circular anisotropy, and stability. The linear and circular degrees contain equal weights, while the contribution from stability is reduced by factor $(L-2)/L$. For a factoring fully orbitalized state $Q(\pmb{\xi})=1$ and $Q_\mathrm{stab}(\pmb{\xi})=1$ and $Q_\mathrm{lin}^2(\pmb{\xi})+Q_\mathrm{circ}^2(\pmb{\xi})=1$, such that the relation is identically satisfied. Also, the limiting linear and circular orbitalization states correspond, respectively, to $[Q_\mathrm{lin}(\pmb{\xi}),Q_\mathrm{circ}(\pmb{\xi}),Q_\mathrm{stab}(\pmb{\xi})]=[1,0,1]$ and $[Q_\mathrm{lin}(\pmb{\xi}),Q_\mathrm{circ}(\pmb{\xi}),Q_\mathrm{stab}(\pmb{\xi})]=[0,1,1]$, with $Q(\pmb{\xi})=1$ in both cases. In addition, $[Q_\mathrm{lin}(\pmb{\xi}),Q_\mathrm{circ}(\pmb{\xi}),Q_\mathrm{stab}(\pmb{\xi})]=[0,0,1]$ implies a perfectly planar state that is completely isotropic within its plane, having $Q^2(\pmb{\xi})=(L-2)(L-1)^{-1}/2$. This case is analogous to a 2D unpolarized state within 3D polarization formalism \cite{Setala2002}.

For $L=3$ the form above reduces to 
\begin{align}
    Q^2(\pmb{\xi})=\frac{3}{4}Q_\mathrm{ell}^2(\pmb{\xi})
    +\frac{1}{4}Q_\mathrm{stab}^2(\pmb{\xi}),
\end{align}
while for $L=2$ we obtain
\begin{align}
Q(\pmb{\xi})=Q_\mathrm{ell}(\pmb{\xi}).
\end{align}
We notice that both results are formally consistent with the corresponding expressions in 3D \cite{Gil2016,Gil2017} and 2D \cite{Gil} polarization theory.

\section{Degrees of linear and circular orbitalization}

In this section we will derive the degrees quantifying the weights of linearly and circularly orbitalized states in the fully orbitalized part of the beam. We note a priori that these degrees are based on a different set of eigenmodes than those  derived in Sec. 4, and cannot be directly deduced from them.

The completely orbitalized contribution to the OM is given by \cite{OErandom}
\begin{align}\label{Oorb}
    \lowarrow{O}_\mathrm{o}(\pmb{\xi})=[\lambda_1(\pmb{\xi})-\lambda_2(\pmb{\xi})]\vec{c}_1(\pmb{\xi})\vec{c}_1^\dagger(\pmb{\xi}).
\end{align}
The associated geometric object is the OE \cite{OE}, as defined for random beams in \cite{OErandom}. Notice that the plane of the OE, specified by $\mathrm{span}[\vec{c}_1'(\pmb{\xi}),\vec{c}_1''(\pmb{\xi})]$, is in general different from the principal plane $\mathbf{A}_{12}(\pmb{\xi})$ defined by $\lowarrow{R}(\pmb{\xi})$. In addition, the Degree of Orbitalization (DO) \cite{OErandom}
\begin{align}
    \mathcal{O}(\pmb{\xi})=\frac{\|\lowarrow{O}_\mathrm{o}(\pmb{\xi})\|_1}{\|\lowarrow{O}(\pmb{\xi})\|_1}=\frac{\lambda_1(\pmb{\xi})-\lambda_2(\pmb{\xi})}{\sum_{n=1}^L\lambda_n(\pmb{\xi})}\label{DO}
\end{align}
quantifies the relative weight of the completely orbitalized part within the total OM. The DO reaches its maximum $\mathcal{O}(\pmb{\xi})=1$ when $\lowarrow{O}(\pmb{\xi})=\lowarrow{O}_\mathrm{o}(\pmb{\xi})$, corresponding to a completely orbitalized beam. The minimum $\mathcal{O}(\pmb{\xi})=0$ is attained for $\lambda_1(\pmb{\xi})=\lambda_2(\pmb{\xi})$, in which case $\lowarrow{O}_\mathrm{o}(\pmb{\xi})$ vanishes.  Matrix $\lowarrow{O}_\mathrm{o}(\pmb{\xi})$ can be decomposed into its real and imaginary parts as $\lowarrow{O}_\mathrm{o}(\pmb{\xi})=\lowarrow{R}_\mathrm{o}(\pmb{\xi})+i\lowarrow{I}_\mathrm{o}(\pmb{\xi})$. We use Eq.~(\ref{cn}) to express the real and imaginary parts of $\lowarrow{O}_\mathrm{o}(\pmb{\xi})$ as
\begin{align}
    \lowarrow{R}_\mathrm{o}(\pmb{\xi})&=\Delta\lambda_{12}(\pmb{\xi})[\vec{c}_1'(\pmb{\xi})\vec{c}_1'^T(\pmb{\xi})+\vec{c}_1''(\pmb{\xi})\vec{c}_1''^T(\pmb{\xi})], \label{Rorb}\\
    \lowarrow{I}_\mathrm{o}(\pmb{\xi})&=\Delta\lambda_{12}(\pmb{\xi})[\vec{c}_1''(\pmb{\xi})\vec{c}_1'^T-\vec{c}_1'(\pmb{\xi})\vec{c}_1''^T(\pmb{\xi})], \label{Iorb}
\end{align}
where $\Delta\lambda_{12}(\pmb{\xi})=\lambda_1(\pmb{\xi})-\lambda_2(\pmb{\xi})$. We immediately see that both $\lowarrow{R}_\mathrm{o}(\pmb{\xi})$ and $\lowarrow{I}_\mathrm{o}(\pmb{\xi})$ are sums of two outer products, hence $\mathrm{rank}[\lowarrow{R}_\mathrm{o}(\pmb{\xi})]\leq 2$ and $\mathrm{rank}[\lowarrow{I}_\mathrm{o}(\pmb{\xi})]\leq 2$.

Denote the non-zero eigenvalues of $\lowarrow{R}_\mathrm{o}(\pmb{\xi})$ with $\nu_1(\pmb{\xi})\geq \nu_2(\pmb{\xi})\geq 0$, and those of $\lowarrow{I}_\mathrm{o}(\pmb{\xi})$ with $\pm i\kappa(\pmb{\xi})$, $\kappa(\pmb{\xi})\in\mathbf{R}$, respectively. Then, the Degree of Linear Orbitalization (DLO) and the Degree of Circular Orbitalization (DCO) can be defined as
\begin{align}
    \mathcal{O}_\mathrm{lin}(\pmb{\xi})&=\frac{\nu_1(\pmb{\xi})-\nu_2(\pmb{\xi})}{\sum_{n=1}^L \lambda_n(\pmb{\xi})}, \label{Olin} \\
    \mathcal{O}_\mathrm{circ}(\pmb{\xi})&=\frac{2|\kappa_1(\pmb{\xi})|}{\sum_{n=1}^L \lambda_n(\pmb{\xi})}. \label{Ocirc}
\end{align}
To obtain interpretation for $\mathcal{O}_\mathrm{lin}(\pmb{\xi})$ and $\mathcal{O}_\mathrm{circ}(\pmb{\xi})$, we next analyze the eigenvalues of $\lowarrow{R}_\mathrm{o}(\pmb{\xi})$ and $\lowarrow{I}_\mathrm{o}(\pmb{\xi})$.
Being at most rank-two matrix, $\lowarrow{R}_\mathrm{o}(\pmb{\xi})$ has at most two non-zero eigenvalues, $\nu_1(\pmb{\xi})$ and $\nu_2(\pmb{\xi})$. We notice from Eq.~(\ref{Rorb}) that 
\begin{align}
    \lowarrow{R}_\mathrm{o}(\pmb{\xi})=\Delta\lambda_{12}(\pmb{\xi})\lowarrow{c}_1(\pmb{\xi})\lowarrow{c}_1^T(\pmb{\xi}),
\end{align}
where $\lowarrow{c}_1(\pmb{\xi})=[\vec{c}_1'(\pmb{\xi}), \vec{c}_1''(\pmb{\xi})]$. This implies that $\lowarrow{R}_\mathrm{o}(\pmb{\xi})$ is a Gram matrix of $\sqrt{\Delta\lambda_{12}(\pmb{\xi})}\lowarrow{c}_1(\pmb{\xi})$. Since the two Gram matrices of the same matrix share the same non-zero eigenvalues (see \cite{Horn}, Th.~2.6.3), $\nu_1(\pmb{\xi})$ and $\nu_2(\pmb{\xi})$ coincide with the eigenvalues of the $2\times 2$ symmetric matrix $\lowarrow{R}_\mathrm{o}'(\pmb{\xi})=\Delta\lambda_{12}(\pmb{\xi})\lowarrow{c}_1^T(\pmb{\xi})\lowarrow{c}_1(\pmb{\xi})$. Consequently, they are given by
\begin{align}
    \nu_n(\pmb{\xi})=&\frac{1}{2}\bigg\{\mathrm{tr}[\lowarrow{R}_\mathrm{o}'(\pmb{\xi})]\nonumber\\&+(-1)^{n-1}\sqrt{\mathrm{tr}^2[\lowarrow{R}_\mathrm{o}'(\pmb{\xi})]-4\det[\lowarrow{R}_\mathrm{o}'(\pmb{\xi})]}]\bigg\}
\end{align}
for $n=1,2$. After straightforward calculations, we obtain
\begin{align}
    \nu_n(\pmb{\xi})=\frac{\Delta\lambda_{12}(\pmb{\xi})}{2}[1+(-1)^{n-1}|\vec{c}_1^T(\pmb{\xi})\vec{c}_1(\pmb{\xi})|]. \label{nu}
\end{align}
Using Eq.~(\ref{cn}), we find that
\begin{align}
    \vec{c}_1^T(\pmb{\xi})\vec{c}_1(\pmb{\xi})=|\vec{c}_1'(\pmb{\xi})|^2-|\vec{c}_1''(\pmb{\xi})|^2+2 i [\vec{c}_1'(\pmb{\xi})\cdot\vec{c}_1''(\pmb{\xi})].
\end{align}
Taking the squared modulus of the above and noting that $|\vec{c}_1(\pmb{\xi})|=1$ gives
\begin{align}
    |\vec{c}_1^T(\pmb{\xi})\vec{c}_1(\pmb{\xi})|^2=1-4\{|\vec{c}_1'(\pmb{\xi})|^2|\vec{c}_1''(\pmb{\xi})|^2-[\vec{c}_1'(\pmb{\xi})\cdot\vec{c}_1''(\pmb{\xi})]^2\}. \label{c1-abs-rel}
\end{align}
The Cauchy--Schwarz inequality \cite{Mandel} states: 
\begin{align}
    [\vec{c}_1'(\pmb{\xi})\cdot\vec{c}_1''(\pmb{\xi})]^2\leq |\vec{c}_1'(\pmb{\xi})|^2|\vec{c}_1''(\pmb{\xi})|^2.
\end{align}
In addition, expressing $\vec{c}_1^T(\pmb{\xi})\vec{c}_1(\pmb{\xi})=\sum_{m=1}^L c_{1m}^2(\pmb{\xi})$, where $c_{1m}^2(\pmb{\xi})$ are the components of $\vec{c}_1(\pmb{\xi})$, and using triangle equality
\begin{align}
    \left|\sum_{m=1}^L c_{1m}^2(\pmb{\xi})\right|^2\leq \sum_{m=1}^L |c_{1m}(\pmb{\xi})|^2=|\vec{c}_1(\pmb{\xi})|^2,
\end{align}
we get
\begin{align}\label{c1-triangle}
    0\leq|\vec{c}_1^T(\pmb{\xi})\vec{c}_1(\pmb{\xi})|\leq 1.
\end{align}
Here $|\vec{c}_1^T(\pmb{\xi})\vec{c}_1(\pmb{\xi})|=0$ when $|\vec{c}_1'(\pmb{\xi})|=|\vec{c}_1''(\pmb{\xi})|$ and $\vec{c}_1'(\pmb{\xi})\cdot\vec{c}_1''(\pmb{\xi})=0$, such that real and imaginary parts of $\vec{c}_1(\pmb{\xi})$ have equal strengths and are orthogonal. This implies that $\vec{c}_1(\pmb{\xi})$, being expressible as a sum of two orthogonal linear states with a mutual phase lag of $\pi/2$, represents a circular orbitalization state. On the other hand, Eq.~(\ref{c1-abs-rel}) shows that $|\vec{c}_1^T(\pmb{\xi})\vec{c}_1(\pmb{\xi})|=1$ when $\vec{c}_1'(\pmb{\xi})$ and $\vec{c}_1''(\pmb{\xi})$ are linearly dependent. Hence, $\vec{c}_1''(\pmb{\xi})=\alpha\vec{c}_1'(\pmb{\xi})$, $\alpha\in\mathbb{R}$, such that $\vec{c}_1(\pmb{\xi})=(1+i\alpha)\vec{c}_1'(\pmb{\xi})$ represents a linear orbitalization state. It then follows from Eq.~(\ref{nu}) that the cases $\nu_2(\pmb{\xi})=0$ and $\nu_1(\pmb{\xi})=\nu_2(\pmb{\xi})$ correspond to $\lowarrow{O}_\mathrm{o}(\pmb{\xi})$ representing a linearly and circularly orbitalized state, respectively. 

Comparing the findings above with Eq.~(\ref{Olin}), we see that $\mathcal{O}_\mathrm{lin}(\pmb{\xi})$ vanishes when there is no linear orbitalization associated with $\lowarrow{O}_\mathrm{o}(\pmb{\xi})$. On the other hand, $\mathcal{O}_\mathrm{lin}(\pmb{\xi})=\mathcal{O}(\pmb{\xi})$ when $\lowarrow{O}_\mathrm{o}(\pmb{\xi})$ represents a completely linear orbitalization state. In addition, $\mathcal{O}_\mathrm{lin}(\pmb{\xi})=1$ is reached when $\nu_2(\pmb{\xi})=\lambda_2(\pmb{\xi})=0$, such that the full OM is $\lowarrow{O}(\pmb{\xi})=\lambda_1(\pmb{\xi})\vec{c}_1(\pmb{\xi})\vec{c}_1^\dagger(\pmb{\xi})$ with $\vec{c}_1(\pmb{\xi})$ representing a linear state. In other words, $\mathcal{O}_\mathrm{lin}(\pmb{\xi})$ attains its maximum for a completely linearly orbitalized beam.

Eigenvalues of $\lowarrow{I}_\mathrm{o}(\pmb{\xi})$ can be found as follows. Since $\lowarrow{I}_\mathrm{o}(\pmb{\xi})$ is a skew-symmetric matrix of at most rank-two, it has at most one conjugate pair of non-zero imaginary eigenvalues, $\pm i \kappa(\pmb{\xi})$. In addition, $\lowarrow{I}_\mathrm{o}(\pmb{\xi})$ is normal, $\lowarrow{I}_\mathrm{o}^T(\pmb{\xi})\lowarrow{I}_\mathrm{o}(\pmb{\xi})=\lowarrow{I}_\mathrm{o}(\pmb{\xi})\lowarrow{I}_\mathrm{o}^T(\pmb{\xi})$, and therefore its squared Schatten 2-norm is given by the sum of the squared magnitudes of its eigenvalues (\cite{Horn}, Th. 2.5.3.):
\begin{align}
    \|\lowarrow{I}_\mathrm{o}(\pmb{\xi})\|_2^2= 2\kappa^2(\pmb{\xi}). \label{Iorb-norm}
\end{align}
Also, using Eqs.~(\ref{Schatten}) and (\ref{Iorb}), we obtain
\begin{align}
    \|\lowarrow{I}_\mathrm{o}(\pmb{\xi})\|_2^2&=-\mathrm{tr}[\lowarrow{I}^2_\mathrm{o}(\pmb{\xi})]\nonumber\\
    &=\frac{\Delta\lambda_{12}(\pmb{\xi})}{2}[1-|\vec{c}_1^T(\pmb{\xi})\vec{c}_1(\pmb{\xi})|^2]. \label{Iorb-norm2}
\end{align}
Combining Eqs.~(\ref{Iorb-norm}) and (\ref{Iorb-norm2}) gives
\begin{align}\label{kappa}
    |\kappa(\pmb{\xi})|=\frac{\Delta\lambda_{12}(\pmb{\xi})}{2}\sqrt{1-|\vec{c}_1^T(\pmb{\xi})\vec{c}_1(\pmb{\xi})|^2}.
\end{align}
Above, $\kappa(\pmb{\xi})=0$ when $|\vec{c}_1^T(\pmb{\xi})\vec{c}_1(\pmb{\xi})|=1$, whereas the maximum $|\kappa(\pmb{\xi})|=\Delta\lambda_{12}(\pmb{\xi})/2$ is obtained when $|\vec{c}_1^T(\pmb{\xi})\vec{c}_1(\pmb{\xi})|=0$. 

In view of Eqs.~(\ref{Ocirc}), (\ref{kappa}), and the discussion below Eq.~(\ref{c1-triangle}), we may conclude that $\mathcal{O}_\mathrm{circ}(\pmb{\xi})=0$ when there is no circular orbitalization contained in $\lowarrow{O}_\mathrm{o}(\pmb{\xi})$, while $\mathcal{O}_\mathrm{circ}(\pmb{\xi})=\mathcal{O}(\pmb{\xi})$ when $\lowarrow{O}_\mathrm{o}(\pmb{\xi})$ represents a completely circularly orbitalized state. Furthermore, $\mathcal{O}_\mathrm{circ}(\pmb{\xi})=1$ is reached for $\lambda_2(\pmb{\xi})=0$ and $|\kappa(\pmb{\xi})|=\lambda_1(\pmb{\xi})/2$, i.e., when $\lowarrow{O}(\pmb{\xi})=\lowarrow{O}_\mathrm{o}(\pmb{\xi})$ and the whole beam is completely circularly orbitalized.

Finally, we see by combining Eqs.~(\ref{DO}), (\ref{Olin}), (\ref{Ocirc}), (\ref{nu}), and (\ref{kappa}) that
\begin{align}\label{O-rel}
\mathcal{O}_\mathrm{lin}^2(\pmb{\xi})+\mathcal{O}_\mathrm{circ}^2(\pmb{\xi})=\mathcal{O}^2(\pmb{\xi}).
\end{align}
We may then summarize as follows. When $\mathcal{O}(\pmb{\xi})=0$, the completely orbitalized part vanishes, and consequently $\mathcal{O}_\mathrm{lin}(\pmb{\xi})=\mathcal{O}_\mathrm{circ}(\pmb{\xi})=0$. For $\mathcal{O}(\pmb{\xi})>0$, if $\mathcal{O}_\mathrm{lin}(\pmb{\xi})=\mathcal{O}(\pmb{\xi})$, the state associated with $\lowarrow{O}_\mathrm{o}(\pmb{\xi})$ is linearly orbitalized, and consequently $\mathcal{O}_\mathrm{circ}(\pmb{\xi})=0$. On the other hand, $\mathcal{O}_\mathrm{circ}(\pmb{\xi})=\mathcal{O}(\pmb{\xi})$ implies that  $\lowarrow{O}_\mathrm{o}(\pmb{\xi})$ represents a circularly orbitalized state, with $\mathcal{O}_\mathrm{lin}(\pmb{\xi})=0$. The absolute maximum values are reached in the following cases: $\mathcal{O}_\mathrm{lin}(\pmb{\xi})=1$ [$\mathcal{O}_\mathrm{circ}(\pmb{\xi})=1$] when the whole beam, rather than only its completely orbitalized part, is linearly (circularly) orbitalized. In this case $\mathcal{O}(\pmb{\xi})=1$, and $\mathcal{O}_\mathrm{lin}(\pmb{\xi})$ and $\mathcal{O}_\mathrm{circ}(\pmb{\xi})$ coincide with $Q_\mathrm{lin}(\pmb{\xi})$ and $Q_\mathrm{circ}(\pmb{\xi})$, respectively.

\section{Concluding remarks}

Thus, we have established a comprehensive framework for characterizing the correlation geometry and topology of scalar random beams endowed with multiple OAM modes. The framework utilizes the OM of the beam, defined at a given cross-section and radius, and decouples the underlying linear and circular orbitalization information encoded in its real and imaginary parts, respectively. While the real part of the OM is shown to define the beam’s correlation geometry via an $L$D orbitalization ellipsoid, its imaginary part possesses a Darboux canonical form that enables association of the beam’s correlation topology with an $N$D orbitalization torus, where $N\leq L/2$. The distribution of all eigenvalues and the orientation of all eigenvectors of the OM, as well as those of its real and imaginary parts, permit the characterization of the correlation structure among the OAM modes of the beam through the degrees of 
linear and circular orbital anisotropy and stability, which contribute to the overall degree of orbital anisotropy. In addition, isolating the fully orbitalized part of the OM makes it possible to determine the degrees of linear and circular orbitalization associated with it, both contributing to the degree of orbitalization. 

Collectively, the revealed orbitalization ellipsoid and torus, along with the associated anisotropy, stability, and orbitalization measures, establish a unified description of correlations in scalar, wide-sense stationary, structured light beams. Moreover, by placing orbital and polarization correlation theories within a common conceptual setting, our framework clarifies both the parallels and distinctions between their respective features and predicts novel classes of correlation structures 
arising from the high-dimensional nature of the OM, beyond those accessible in polarization optics.

\qquad 

\section*{Acknowledgments}
This work was supported by the Finnish Cultural Foundation.

\appendix

\section{Alternative representation of matrices $\lowarrow{R}(\pmb{\xi})$ and $\lowarrow{I}(\pmb{\xi})$.} \label{appA}
For the real and imaginary parts of the OM one can also write 
\begin{align}
&\lowarrow{R}(\pmb{\xi})
= \langle \vec{E}^* (\pmb{\xi}) \odot\vec{E} (\pmb{\xi})\rangle/2, \\& \lowarrow{I}(\pmb{\xi})= -i\langle \vec{E}^*(\pmb{\xi})\wedge \vec{E} (\pmb{\xi})\rangle/2,
\end{align}
where $\odot$ and $\wedge$ are Jordan and wedge products, respectively: $\vec{X}\odot \vec{Y}=\vec{X} \otimes \vec{Y}+ \vec{Y} \otimes \vec{X}$ and $\vec{X}\wedge \vec{Y}=\vec{X} \otimes \vec{Y}- \vec{Y} \otimes \vec{X}$, $\otimes$ being tensor (outer) product \cite{greub}. In terms of individual elements the two products are:
\begin{align}\label{wedge}
&\langle \vec{E}^* (\pmb{\xi}) \odot\vec{E} (\pmb{\xi})\rangle_{lm}=\langle E_l^\ast(\pmb{\xi}) E_m(\pmb{\xi})\rangle+ \langle E_m^\ast(\pmb{\xi}) E_l(\pmb{\xi})\rangle, \\& 
\langle \vec{E}^\ast(\pmb{\xi}) \wedge \vec{E}(\pmb{\xi})\rangle_{lm}=\langle E_l^\ast(\pmb{\xi}) E_m(\pmb{\xi})\rangle- \langle E_m^\ast(\pmb{\xi}) E_l(\pmb{\xi})\rangle.
\end{align}

\section{Proof of the statement that $\text{rank}[\lowarrow{I}(\pmb{\xi})]= 2$ or 0 for correlated beams. } \label{appB}
Since matrix $\lowarrow{I}(\pmb{\xi})$ in \eqref{Idet} is the sum of two outer products, its every column is a linear combination of $\vec{E}^R(\pmb{\xi})$ and $\vec{E}^I(\pmb{\xi})$. This implies that the column space of $\lowarrow{I}(\pmb{\xi})$ is contained in $\mathrm{span}\{\vec{E}^R(\pmb{\xi}),\vec{E}^I(\pmb{\xi})\}$ and therefore $\mathrm{rank}[\lowarrow{I}(\pmb{\xi})]\leq 2$. Further, if  $\vec{E}^R(\pmb{\xi})$ and $\vec{E}^I(\pmb{\xi})$ are linearly dependent, then the two outer products cancel, giving $\lowarrow{I}(\pmb{\xi})=0$ and hence $\mathrm{rank}[\lowarrow{I}(\pmb{\xi})]=0$. On the contrary, if they are linearly independent, then both vectors contribute independently to the column space, making it  exactly 2D, leading to $\mathrm{rank}[\lowarrow{I}(\pmb{\xi})]=2$. Thus, $\mathrm{rank}[\lowarrow{I}(\pmb{\xi})]\in\{0,2\}$.

\section{Projection of invariant torus $\textbf{T}^2(\pmb{\xi})$ to torus in 3D.} \label{appC}
For $N=2$, the system evolves on $\textbf{T}^2(\pmb{\xi})\subset\mathbb{R}^4$. Its embedding into the unit 3-sphere $S^3\subset\mathbb{R}^4$ is given by 
\begin{align}
\begin{split}
&\tilde{x}_1(\pmb{\xi},s)=R_1(\pmb{\xi}) \cos[\Psi_1(\pmb{\xi})-\eta_1(\pmb{\xi})s]/R_{12}(\pmb{\xi}), \\& \tilde{x}_2(\pmb{\xi},s)=R_1(\pmb{\xi}) \sin[\Psi_1(\pmb{\xi})-\eta_1(\pmb{\xi})s]/R_{12}(\pmb{\xi}),  \\& \tilde{x}_3(\pmb{\xi},s)=R_2(\pmb{\xi})\cos[\Psi_2(\pmb{\xi})-\eta_2(\pmb{\xi})s]/R_{12}(\pmb{\xi}), \\& \tilde{x}_4(\pmb{\xi},s)=R_2(\pmb{\xi}) \sin[\Psi_2(\pmb{\xi})-\eta_2(\pmb{\xi})s]/R_{12}(\pmb{\xi}),  
\end{split}
\end{align}
where $R_n(\pmb{\xi})$ and $\Psi_n(\pmb{\xi})$, $n=1,2$, are given in \eqref{Rn} and \eqref{Psin}, respectively, $R_{12}(\pmb{\xi})=\sqrt{R_1^2(\pmb{\xi})+R_2^2(\pmb{\xi})}$, and  symbol tilde is used to distinguish coordinates normalized by $R_{12}(\pmb{\xi})$ from those given in \eqref{flow}. The resulting curve lies on a Clifford-type torus in $S^3$. A 3D visualization is obtained via stereographic projection from $S^3$ to $\mathbb{R}^3$, on omitting dependence on $\pmb{\xi}$ and $s$:
\begin{align}
(X,Y,Z)=\left(\frac{\tilde{x}_1}{1-\tilde{x}_4},\frac{\tilde{x}_2}{1-\tilde{x}_4},\frac{\tilde{x}_3}{1-\tilde{x}_4}\right),
\end{align}
mapping $\textbf{T}^2(\pmb{\xi})$ to a surface in $\mathbb{R}^3$ and the trajectory to a curve on it.

\section{Decomposition of the OM with real embedding.}  \label{appD}
Transformation (\ref{Ospec}) suggests that the OM may be factored as
\begin{equation}
\lowarrow{O}(\pmb{\xi}) = \lowarrow{Z}(\pmb{\xi})\lowarrow{Z}^{\dagger}(\pmb{\xi}), \quad \lowarrow{Z}(\pmb{\xi}) \in \mathbb{C}^{L\times L},
\end{equation}
with $\lowarrow{Z}(\pmb{\xi})=\lowarrow{C}(\pmb{\xi})\lowarrow{\Lambda}^{1/2}(\pmb{\xi})=\lowarrow{X}(\pmb{\xi})+i\lowarrow{Y}(\pmb{\xi})$, yielding 
\begin{align}
\lowarrow{O}(\pmb{\xi}) = \tilde{Z}(\pmb{\xi})\tilde{Z}^T(\pmb{\xi})+i\tilde{Z}(\pmb{\xi})\lowarrow{J}_L \tilde{Z}^T(\pmb{\xi}), 
\end{align}
where $\tilde{Z}(\pmb{\xi})=[\lowarrow{X}(\pmb{\xi}), \lowarrow{Y}(\pmb{\xi})]\in \mathbb{R}^{L\times 2L}$ is row-based real embedding and
\begin{align}
    \lowarrow{J}_L=\begin{bmatrix}
        0 & -\lowarrow{I}_L \\ \lowarrow{I}_L & 0
    \end{bmatrix},
\end{align}
with $\lowarrow{I}_L$ being the $L\times L$ identity matrix. Element-wise, this representation gives
\begin{align}
&R_{lm}(\pmb{\xi})=\tilde{z}_{l}^T(\pmb{\xi})\tilde{z}_{m}(\pmb{\xi})=\vec{x}_l(\pmb{\xi})\cdot\vec{x}_m(\pmb{\xi})+\vec{y}_l(\pmb{\xi})\cdot\vec{y}_m(\pmb{\xi}), 
\\&
I_{lm}(\pmb{\xi})=\tilde{z}_{l}^T(\pmb{\xi})\lowarrow{J}_L\tilde{z}_{m}(\pmb{\xi})=\vec{y}_l(\pmb{\xi})\cdot\vec{x}_m(\pmb{\xi})-\vec{x}_l(\pmb{\xi})\cdot\vec{y}_m(\pmb{\xi}),
\end{align}
where 
\begin{align}
\begin{split}
\tilde{Z}(\pmb{\xi})&=[\tilde{z}_1(\pmb{\xi}),...,\tilde{z}_L(\pmb{\xi})]^T, \\ 
\lowarrow{X}(\pmb{\xi})&=[\vec{x}_1(\pmb{\xi}),...,\vec{x}_L(\pmb{\xi})]^T, \\
\lowarrow{Y}(\pmb{\xi})&=[\vec{y}_1(\pmb{\xi}),...,\vec{y}_L(\pmb{\xi})]^T.
\end{split}
\end{align}

\section{Bounds of norms $\|\lowarrow{I}(\pmb{\xi})\|_1$ and $\|\lowarrow{I}(\pmb{\xi})\|_2$}\label{appE}

Here we derive the lower and upper bounds of the Schatten 1- and 2-norms of $\lowarrow{I}(\pmb{\xi})$ and inspect their implications for the orbital anisotropy of the beam. 

Revisiting Eq.~(\ref{PT}), the norms 
\begin{align}
\|\lowarrow{I}(\pmb{\xi})\|_1=2\sum_{n=1}^N|\eta_n(\pmb{\xi})|,\quad \|\lowarrow{I}(\pmb{\xi})\|_2=\left[2\sum_{n=1}^N\eta_n^2(\pmb{\xi})\right]^{1/2},    
\end{align}
offer two alternative ways of defining the DCOA. The 1-norm of $\lowarrow{I}(\pmb{\xi})$ adds the magnitudes of $\eta_n(\pmb{\xi})$ linearly, while the 2-norm combines them quadratically. We notice that both norms vanish when $\lowarrow{I}(\pmb{\xi})$ has no non-zero eigenvalues and thus no circular orbitalization is present in the beam. The upper bounds are discussed below.

We start by discussing the 1-norm. Employing the spectral decomposition of $\lowarrow{O}(\pmb{\xi})$, Eq.~(\ref{Ospec2}), and Eq.~(\ref{cn}), we may express the OM as
\begin{align}
    \lowarrow{O}(\pmb{\xi})=\sum_{n=1}^L[\lowarrow{R}_n(\pmb{\xi})+i\lowarrow{I}_n(\pmb{\xi})],
\end{align}
where 
\begin{align}
    \lowarrow{R}_n(\pmb{\xi})&=\lambda_n(\pmb{\xi})[\vec{c}_n'(\pmb{\xi})\vec{c}_n'^T(\pmb{\xi})+\vec{c}_n''(\pmb{\xi})\vec{c}_n''^T(\pmb{\xi})], \\
    \lowarrow{I}_n(\pmb{\xi})&=\lambda_n(\pmb{\xi})[\vec{c}_n''(\pmb{\xi})\vec{c}_n'^T(\pmb{\xi})-\vec{c}_n'(\pmb{\xi})\vec{c}_n''^T].
\end{align}
It is readily seen that 
\begin{align}
    \lowarrow{I}(\pmb{\xi})=\sum_{n=1}^L\lowarrow{I}_n(\pmb{\xi}).
\end{align}
Using triangle inequality, we obtain upper bound for the 1-norm:
\begin{align}\label{I-ineq}
    \|\lowarrow{I}(\pmb{\xi})\|_1\leq \sum_{n=1}^L|\lowarrow{I}_n(\pmb{\xi})\|_1.
\end{align}
To compute $|\lowarrow{I}_n(\pmb{\xi})\|_1=\mathrm{tr}\{[\lowarrow{I}_n^T(\pmb{\xi})\lowarrow{I}_n(\pmb{\xi})]^{1/2}\}$, we write $\lowarrow{I}_n(\pmb{\xi})=\lambda_n(\pmb{\xi})\lowarrow{K}_n(\pmb{\xi})$, where $\lowarrow{K}_n(\pmb{\xi})=\vec{c}_n''(\pmb{\xi})\vec{c}_n'^T(\pmb{\xi})-\vec{c}_n'(\pmb{\xi})\vec{c}_n''^T$. 
Then, straightforward calculations show that
\begin{align}
    [\lowarrow{K}_n^T(\pmb{\xi})\lowarrow{K}_n(\pmb{\xi})]^2=\Omega_n(\pmb{\xi})\lowarrow{K}_n^T(\pmb{\xi})\lowarrow{K}_n(\pmb{\xi}),
\end{align}
where 
\begin{align}
    \Omega_n(\pmb{\xi})&=|\vec{c}_n'(\pmb{\xi})|^2|\vec{c}_n''(\pmb{\xi})|^2-[\vec{c}_n'(\pmb{\xi})\cdot\vec{c}_n''(\pmb{\xi})]^2\nonumber\\
    &=\frac{1}{4}\left[1-|\vec{c}_n^T(\pmb{\xi})\vec{c}_n(\pmb{\xi})|^2\right],
\end{align}
where the property $\vec{c}_n^\dagger(\pmb{\xi})\vec{c}_n(\pmb{\xi})=1$ has been used to obtain the second equality. Consequently, 
\begin{align}
[\lowarrow{K}_n^T(\pmb{\xi})\lowarrow{K}_n(\pmb{\xi})]^{1/2}=\frac{\lowarrow{K}_n^T(\pmb{\xi})\lowarrow{K}_n(\pmb{\xi})}{\Omega_n^{1/2}(\pmb{\xi})},
\end{align}
and thus
\begin{align}
    \mathrm{tr}\{[\lowarrow{K}_n^T(\pmb{\xi})\lowarrow{K}_n(\pmb{\xi})]^{1/2}\}=2\Omega_n^{1/2}(\pmb{\xi}).
\end{align}
This yields
\begin{align}
    \|\lowarrow{I}_n(\pmb{\xi})\|_1=\lambda_n(\pmb{\xi})\sqrt{1-|\vec{c}_n^T(\pmb{\xi})\vec{c}_n(\pmb{\xi})|^2}.
\end{align}
Noting that $|\vec{c}_n^T(\pmb{\xi})\vec{c}_n(\pmb{\xi})|^2\leq 1$, we obtain the inequality
\begin{align}
    \sum_{n=1}^L \lambda_n(\pmb{\xi})\sqrt{1-|\vec{c}_n^T(\pmb{\xi})\vec{c}_n(\pmb{\xi})|^2}\leq \sum_{n=1}^L\lambda_n(\pmb{\xi})
\end{align}
which together with Eqs.~(\ref{PT}) and (\ref{I-ineq}) shows that 
\begin{align}
\|\lowarrow{I}(\pmb{\xi})\|_1\leq \|\lowarrow{O}(\pmb{\xi})\|_1.
\end{align}
Above, equality is achieved when $|\vec{c}_n^T(\pmb{\xi})\vec{c}_n(\pmb{\xi})|=0$, implying that $\vec{c}_n'(\pmb{\xi})\perp\vec{c}_n''(\pmb{\xi})$ and $|\vec{c}_n'(\pmb{\xi})|=|\vec{c}_n''(\pmb{\xi})|=1/\sqrt{2}$ for all $1\leq n\leq L$. Consequently, $\|\lowarrow{I}(\pmb{\xi})\|_1$ reaches its maximum value when every $\vec{c}_n(\pmb{\xi})$ in the spectral decomposition of $\lowarrow{O}(\pmb{\xi})$ represents a circular orbitalization state. However, several mutually orthogonal such factors may be present. Consequently, maximal $\|\lowarrow{I}(\pmb{\xi})\|_1=\|\lowarrow{O}(\pmb{\xi})\|_1$ does not in general imply a factoring OM, i.e., the orbitalization state of the beam is not necessarily circular and thus need not correspond to complete circular anisotropy, or even full anisotropy. 

For the 2-norm, we revisit again Eq.~(\ref{PT}):
\begin{align}
    \|\lowarrow{O}\|_2^2=\|\lowarrow{R}\|_2^2=+\|\lowarrow{I}\|_2^2=\sum_{n=1}^L\lambda_n^2(\pmb{\xi}).
\end{align}
In addition, we notice that 
\begin{align}
    \mathrm{tr}[\lowarrow{O}(\pmb{\xi})\lowarrow{O}^\ast(\pmb{\xi})]&=\|\lowarrow{R}\|_2^2-\|\lowarrow{I}\|_2^2\nonumber\\
    &=\sum_{n=1}^L\sum_{m=1}^L \lambda_n(\pmb{\xi})\lambda_m(\pmb{\xi})|\vec{c}_n^T(\pmb{\xi})\vec{c}_m(\pmb{\xi})|^2,
\end{align}
such that
\begin{align}
    &\|\lowarrow{I}\|_2^2=\frac{1}{2}\left\{\|\lowarrow{O}\|_2^2-\mathrm{tr}[\lowarrow{O}(\pmb{\xi})\lowarrow{O}^\ast(\pmb{\xi})]\right\}\nonumber\\
    &=\frac{1}{2}\left[\sum_{n=1}^L \lambda_n^2(\pmb{\xi})-\sum_{n=1}^L\sum_{m=1}^L \lambda_n(\pmb{\xi})\lambda_m(\pmb{\xi})|\vec{c}_n^T(\pmb{\xi})\vec{c}_m(\pmb{\xi})|^2\right].
\end{align}
The second term on the second line above is non-negative, yielding inequality 
\begin{align}\label{I2-ineq}
    \|\lowarrow{I}\|_2^2\leq \frac{1}{2}\sum_{n=1}^L \lambda_n^2(\pmb{\xi})
\end{align}
In addition,
\begin{align}\label{lambda-ineq}
    \sum_{n=1}^L \lambda_n^2(\pmb{\xi})\leq \Big[\sum_{n=1}^L \lambda_n(\pmb{\xi})\Big]^2=\|\lowarrow{O}\|_2.
\end{align}
Combining Eqs.~(\ref{I2-ineq}) and (\ref{lambda-ineq}), we obtain
\begin{align}\label{I2-ine2}
    \|\lowarrow{I}\|_2^2\leq \frac{1}{2}\|\lowarrow{O}\|_2.
\end{align}
Equality in Eq.~(\ref{I2-ineq}) is reached when, for non-zero $\lambda_n(\pmb{\xi})$ and $\lambda_m(\pmb{\xi})$, we have $|\vec{c}_n^T(\pmb{\xi})\vec{c}_m(\pmb{\xi})|=0$, i.e., also for $n=m$, such that $\vec{c}_n(\pmb{\xi})$ represents a circular orbitalization state as noted above. Then, equality in Eq.~(\ref{lambda-ineq}) requires that $\lambda_n(\pmb{\xi})=0$ for $n\geq 2$, i.e., only the first eigenvalue is non-zero. Thus, the upper bound $\|\lowarrow{I}(\pmb{\xi})\|_2=\|\lowarrow{O}(\pmb{\xi})\|_1/\sqrt{2}$ is obtained when $\lowarrow{O}(\pmb{\xi})=\lambda_1(\pmb{\xi})\vec{c}_1(\pmb{\xi})\vec{c}_1^\dagger(\pmb{\xi})$, where $\vec{c}_1(\pmb{\xi})$ represents a circular orbitalization state. Consequently, maximum of the 2-norm  is attained only when the OM consists of a single factoring completely circular orbitalization state.

\bibliography{sample}

\end{document}